\documentclass[nofootinbib, reprint, preprintnumbers, amsmath, amssymb, amsfonts, aps, pra, superscriptaddress, floatfix]{revtex4-2}
\usepackage[utf8]{inputenc}
\usepackage{mathtools}
\usepackage{graphicx}
\usepackage{dcolumn}
\usepackage{bm}
\usepackage{physics}
\usepackage{color}
\usepackage{braket}
\usepackage[normalem]{ulem}
\usepackage{enumerate}
\usepackage{enumitem}
\usepackage[colorlinks=true,linkcolor=blue,citecolor=blue,urlcolor =blue]{hyperref}
\usepackage{mathrsfs}
\usepackage{bbm}
\usepackage{siunitx}

\usepackage{comment}
\usepackage{lipsum}
\newcommand{\YM}[1]{\textcolor{black}{#1}}
\newcommand{\AdMORI}[1]{\textcolor{black}{#1}}

\begin{document}
\title{Theoretical study of the Spectroscopic measurements of Kerr non-linear resonators with four-body interaction}

\author{Yuichiro Matsuzaki}
\affiliation{Department of Electrical, Electronic, and Communication Engineering, Faculty of Science and Engineering, Chuo university, 1-13-27, Kasuga, Bunkyo-ku, Tokyo 112-8551, Japan}%

 \author{Yuichiro Mori}
\affiliation{Global Research and Development Center for Business by Quantum-AI Technology (G-QuAT), National Institute of Advanced Industrial Science and Technology (AIST), 1-1-1, Umezono, Tsukuba, Ibaraki 305-8568, Japan}%

 \author{Aiko Yamaguchi}
 \affiliation{Secure System Platform Research Laboratories, NEC Corporation, 
1753, Shimonumabe, Kawasaki, Kanagawa 211-0011, Japan}
 \affiliation{NEC-AIST Quantum Technology Cooperative Research Laboratory, National Institute of Advanced Industrial Science and Technology (AIST), Tsukuba, Ibaraki 305-8568, Japan}

\author{Yohei Kawakami}
 \affiliation{Secure System Platform Research Laboratories, NEC Corporation, 
1753, Shimonumabe, Kawasaki, Kanagawa 211-0011, Japan}
 \affiliation{NEC-AIST Quantum Technology Cooperative Research Laboratory, National Institute of Advanced Industrial Science and Technology (AIST), Tsukuba, Ibaraki 305-8568, Japan}

  \author{Tsuyoshi Yamamoto}
 \affiliation{Secure System Platform Research Laboratories, NEC Corporation, 
1753, Shimonumabe, Kawasaki, Kanagawa 211-0011, Japan}
 \affiliation{NEC-AIST Quantum Technology Cooperative Research Laboratory, National Institute of Advanced Industrial Science and Technology (AIST), Tsukuba, Ibaraki 305-8568, Japan}

\date{\today}


\begin{abstract}
Quantum annealing provides a promising way to solve combinational optimization problems where the solutions 
\textcolor{black}{correspond to}
the ground state of the Ising Hamiltonian. We can implement quantum annealing using the Kerr non-linear resonators, with bifurcation phenomena emerging when subjected to a parametric drive. These bifurcated states can function \textcolor{black}{as bases of qubits} qubits. Moreover, integrating four-body interactions between physical qubits enables the establishment of \textcolor{black}{effective all-to-all}
long-range interactions between logical qubits, which is essential for 
\textcolor{black}{practical}
quantum annealing. While theoretical proposals exist for creating four-body interactions within Kerr non-linear resonators, 
there has not been experimental verification through their spectroscopic signatures.
In this paper, we theoretically investigate the spectroscopic measurements of Kerr non-linear resonators featuring four-body interaction. We identify six distinct frequencies exhibiting population changes by employing resonant driving on one resonator and weak driving on another.
Analytical and numerical calculations validate these findings. Our study demonstrates the potential of spectroscopy in characterizing systems with four-body interactions, offering insights for realizing quantum annealing with Kerr parametric oscillators.
\end{abstract}

\maketitle
\section{Introduction}
\label{sec:intro}
Quantum annealing is one of the methods for solving combinational optimization problems\cite{Kadowaki_1998_pre,farhi2000quantum,farhi2001quantum}. Solutions to the combinational optimization problems can be embedded in the ground state of the Ising Hamiltonian \cite{choi2011minor}. By evolving in time to satisfy adiabaticity \cite{childs2001robustness,morita2008mathematical}, quantum annealing allows us to prepare the ground state of the Ising Hamiltonian.
Conventionally, superconducting qubits have been utilized to implement quantum annealing \cite{harris2010experimental,harris2010experimental2}. Remarkably, D-Wave has achieved a significant milestone by pioneering the development of a quantum annealing device composed of thousands of superconducting qubits \cite{johnson2011quantum,king2023quantum}.

The Kerr parametric oscillator (KPO) has attracted attention as an alternative approach for realizing quantum annealing 
\YM{\cite{milburn1991quantum,wielinga1993quantum,meaney2014quantum,goto2016bifurcation,Puri2017_npjq,wang2019quantum,grimm2020stabilization,yamaji2022spectroscopic,yamaji2023correlated,iyama2024observation,yamaguchi2023spectroscopy}}. Bifurcation phenomena occur by applying a parametric drive to a Kerr non-linear resonator (KNR) \cite{milburn1991quantum,wielinga1993quantum,meaney2014quantum}. These bifurcated states are then treated as the 0 or 1 state of a qubit 
\cite{cochrane2000teleportation}. While the KPOs have traditionally been investigated as elements for gate-type quantum computers \cite{cochrane2000teleportation}, recent theoretical studies have shown that they can also be used to find the ground state of an Ising Hamiltonian via the adiabatic process \cite{goto2016bifurcation,Puri2017_npjq}. Moreover, experiments have been conducted in this regard \YM{\cite{yamaji2023correlated}}.
It is worth mentioning that, similar to the superconducting transmon qubit \cite{koch2007charge,schreier2008suppressing}, we can treat the KNR
without a parametric drive as a qubit where the Fock states of $|0\rangle $ and $|1\rangle $ play a role in the qubit while the other Fock states are not involved in the dynamics by using a frequency selectivity due to the Kerr non-linearity.

Long-range interactions typically arise when mapping optimization problems onto the ground state of the Ising Hamiltonian \cite{santoro2002theory}. However, a prevalent challenge in practical quantum devices lies in the limitation of establishing long-range interactions because the physical qubits interact locally by nature \cite{harris2010experimental,harris2010experimental2}. Lechner, Hauke, and Zoller (LHZ) recently proposed a strategy to overcome this challenge \cite{lechner2015quantum}. 
They use $N(N-1)/2$ physical qubits to form $N$ logical qubits.
Consequently, while only nearest-neighbor interactions are present among the physical qubits, effectively creating fully connected interactions between logical qubits is feasible. Nonetheless, this method requires the incorporation of four-body interactions among physical qubits.
Despite theoretical proposals for realizing such four-body interactions between superconducting circuits
\cite{puri2017quantum,zhao2018two,razmkhah2024josephson}, 
there have been no experimental reports yet.
If a system with four-body interaction were realized, it would be essential to characterize the system in the experiment.
For a system with two-body interaction, the avoided crossing phenomenon is commonly observed in spectroscopy~\cite{weber2017coherent}, which helps confirm the existence of the two-body interaction.
On the other hand, to our knowledge, there is no established method to characterize the system with four-body interaction by spectroscopy.


Here, we
theoretically study the spectroscopic measurements of the 
KNRs
with four-body interaction.
In our method, we resonantly drive one KNR while weakly driving another KNR by sweeping the driving frequency, and we measure the population of the excited state of the other KNRs.
We show that changes in the population can be observed at six frequencies, which is supported by analytical and numerical calculations. 
When we plot the population against the driving frequency, the change in the population can be observed as peaks.
These peaks let us know the 
 energy level structure of the system,
which is essential to realize quantum annealing with KPOs.

The remainder of this paper is organized as follows.
In Sec. II, we explain the Hamiltonian between the KNRs. 
In Sec. III, we show our analytical results.
In Sec. IV, we introduce our numerical results by taking into account decoherence. 
Finally, in Sec. V, we conclude our discussion.

\section{Hamiltonian}
\begin{figure}
    \centering
    \includegraphics[width = 9cm]{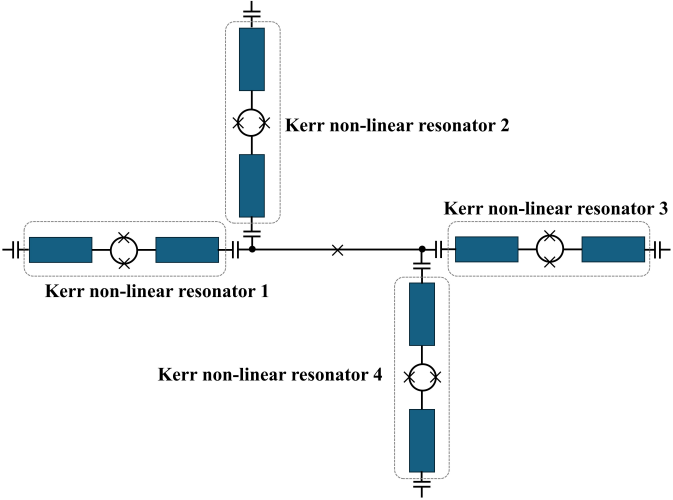}
    \caption{Four Kerr non-linear resonators are capacitively coupled with a Josephson junction mode. The resonant frequencies of the Kerr non-linear resonators are far detuned from that of the Josephson junction mode. The non-linearity of the Josephson junction induces an effective four-body interaction between the Kerr non-linear resonators.
    }
    \label{circuit}
\end{figure}

We can realize the following Hamiltonian by using a superconducting circuit \cite{Puri2017_npjq} as shown in Fig. \ref{circuit}.
In a rotating frame defined by \YM{$U=e^{it\sum _{j=1}^4 \omega^{(j)}_{\rm{rot}}  \hat{a}^{\dagger}_j\hat{a}_j}$}, by using the rotating wave approximation \cite{Puri2017_npjq}, we obtain 
\YM{
\begin{eqnarray}
    &&H=\sum _{j=1}^4 \Big{(}
    (\Tilde{\omega} _j-\omega^{(j)}_{\rm{rot}} ) \hat{a}^{\dagger}_j\hat{a}_j+ K_j \hat{a}^{\dagger}_j\hat{a}_j\hat{a}^{\dagger}_j
    \hat{a}_j \nonumber \\
    &+&
    \lambda _j  (\hat{a}_je^{-i(\omega^{(j)}_{\rm{rot}}-\omega '_j)t} + \hat{a}^{\dagger}_j
    e^{i(\omega^{(j)}_{\rm{rot}}-\omega '_j)t})\Big{)}\nonumber \\
    &+&g(\hat{a}^{\dagger}_1 
    \hat{a}^{\dagger}_2  \hat{a}_3\hat{a}_4+\hat{a}_1 
    \hat{a}_2  \hat{a}^{\dagger}_3\hat{a}^{\dagger}_4)\nonumber \\
    &+&\sum _{i<j} J_{ij} \hat{a}^{\dagger}_i \hat{a}_i \hat{a}^{\dagger}_j \hat{a}_j,
\end{eqnarray}}where $\Tilde{\omega}_j$
\YM{($\omega^{(j)}_{\rm{rot}}$)}
is the resonance \YM{(rotating-frame)} frequency of the $j$-th KNR,
$K_j$ is the Kerr coefficient, $\lambda_j$ is the  driving strength, $\omega_j'$ is the frequency of the driving field, $g$ is the strength of the four-body interaction, and
$J_{ij}$ is the coupling strength of the two-body dispersive interaction between
the $i$-th KNR and the $j$-th KNR.
\YM{Here, we assume
$\omega ^{(1)}_{\rm{rot}}+\omega ^{(2)}_{\rm{rot}}=\omega ^{(3)}_{\rm{rot}}+\omega ^{(4)}_{\rm{rot}}$ and $\omega ^{(i)}_{\rm{rot}}\neq \omega ^{(j)}_{\rm{rot}}$ for $i\neq j$.}

We can simplify the Hamiltonian as follows.
We can treat the KNR as a qubit
without a parametric drive where the Fock states of $|0\rangle $ and $|1\rangle $ play a role of the qubit.
In this case, the other Fock states are not involved in the dynamics 
because their frequencies are different from the qubit transition frequency
due to the Kerr non-linearity.
In this case, we can consider a subspace spanned by $|0\rangle$ and $|1\rangle$, and we obtain the following Hamiltonian:
\textcolor{black}{
\begin{eqnarray}
H&=&\sum _{j=1}^4( \frac{(\tilde{\omega} _j - \omega^{(j)}_{\rm{rot}})}{2} \hat{\sigma}_z^{(j)}
+ \lambda_j (\hat{\sigma}_-^{(j)}e^{-i(\omega^{(j)}_{\rm{rot}} -\omega_j')t}+\hat{\sigma}_+^{(j)}e^{i(\omega^{(j)}_{\rm{rot}} -\omega_j')t})
)\nonumber \\
&+&g(\hat{\sigma}_+^{(1)}\hat{\sigma}_+^{(2)} \hat{\sigma}_-^{(3)}\hat{\sigma}_-^{(4)}+ \hat{\sigma}_-^{(1)}\hat{\sigma}_-^{(2)}
\hat{\sigma}_+^{(3)}
\hat{\sigma}_+^{(4)})\nonumber \\
&+&\sum _{i<j}J_{ij} \frac{\hat{1}+\hat{\sigma}_z ^{(i)}}{2}\frac{\hat{1}+\hat{\sigma}_z ^{(j)}}{2},
\end{eqnarray}
}
Here, $\hat{\sigma}_z =|1\rangle \langle 1|-|0\rangle \langle 0|$ denotes the Pauli matrix, while $\hat{\sigma}_{+}=|1\rangle \langle 0|$ and $\hat{\sigma}_{-}=|0\rangle \langle 1|$ denote the ladder operators.
By using 
shifted frequencies 
 \YM{$\omega_j=\tilde{\omega}_j + \frac{1}{2}\sum _{i< j} J_{ij}+\frac{1}{2}\sum _{i< j} J_{ji}$}
we can rewrite the Hamiltonian as
\begin{eqnarray}
H&=&\sum _{j=1}^4\Big{(} \frac{(\omega _j - \omega^{(j)}_{\rm{rot}})}{2} \hat{\sigma}_z
+ \lambda_j (\hat{\sigma}_-^{(j)}e^{-i(\omega^{(j)}_{\rm{rot}} -\omega_j')t}+\hat{\sigma}_+^{(j)}e^{i(\omega^{(j)}_{\rm{rot}} -\omega_j')t})
\Big{)}\nonumber \\
&+&g(\hat{\sigma}_+^{(1)}\hat{\sigma}_+^{(2)} \hat{\sigma}_-^{(3)}\hat{\sigma}_-^{(4)}+ \hat{\sigma}_-^{(1)}\hat{\sigma}_-^{(2)}
\hat{\sigma}_+^{(3)}
\hat{\sigma}_+^{(4)})\nonumber \\
&+&\sum _{i<j}\frac{J_{ij}}{4}\hat{\sigma}_z ^{(i)}\hat{\sigma}_z ^{(j)}.
\end{eqnarray}
Throughout the paper, we set
$\lambda_1 =\lambda/2$,
$\lambda_3=\lambda_4=0$,  
 \textcolor{black}{$\omega^{(1)}_{\rm{rot}}=\omega_1'=\omega_1$}, 
\YM{$\omega_1+\omega_2=\omega_3+\omega_4$, and $\omega_i\neq \omega _j$ for $i \neq j$.}
\YM{We drive the first qubit to induce transitions between the ground and excited states. On the other hand, we weakly drive the second qubit to probe the system.}
We use the notation of 
$|1\rangle =|\!\uparrow \rangle $ 
and $|0\rangle =|\!\downarrow \rangle $.
\textcolor{black}{We can tune 
$\Tilde{\omega} _j$
by changing the magnetic flux penetrating the SQUID structure embedded in the resonator \cite{yamaji2022spectroscopic}.}
In this case, the Hamiltonian is written as follows:
\YM{
\begin{eqnarray}
    H&=&H_0 + H_{\rm{drive}}(t),\label{simpleh} \\
    H_0&=&  (\sum _{j=1}^4 \frac{\omega _j - \omega^{(j)}_{\rm{rot}}}{2} \hat{\sigma}_z^{(j)})+
    \frac{\lambda}{2} \hat{\sigma}_x^{(1)}
    \nonumber \\
&+&g(\hat{\sigma}_+^{(1)}\hat{\sigma}_+^{(2)} \hat{\sigma}_-^{(3)}\hat{\sigma}_-^{(4)}+ \hat{\sigma}_-^{(1)}\hat{\sigma}_-^{(2)}
\hat{\sigma}_+^{(3)}
\hat{\sigma}_+^{(4)})\nonumber \\
&+&\sum _{i<j}\frac{J_{ij}}{4}\hat{\sigma}_z ^{(i)}\hat{\sigma}_z ^{(j)}
,
\label{mainhamiltonian}
 \\
H_{\rm{drive}}(t)&=&
\lambda_2 (\hat{\sigma}_-^{(2)}e^{i(\omega_2' -\omega^{(2)}_{\rm{rot}})t}+\hat{\sigma}_+^{(2)}e^{-i(\omega_2' -\omega^{(2)}_{\rm{rot}})t}).
\end{eqnarray}}

\section{Analytical results}
\begin{figure}
    \centering
    \includegraphics[width = 9cm]{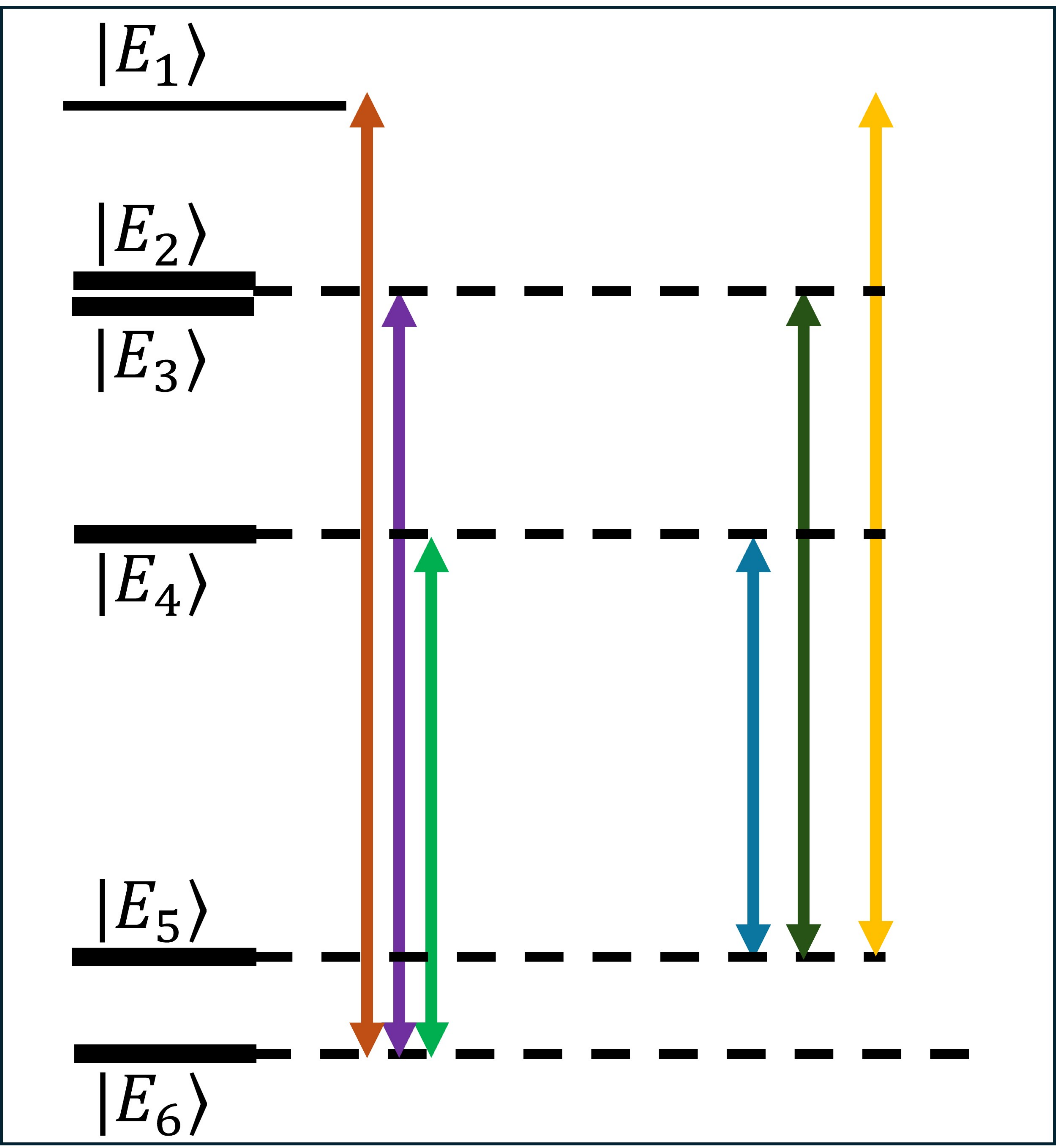}
    \caption{Energy diagram of our system. Six transitions should be observed.
    Here, we assume the four-body interaction is much larger than the two-body interaction \textcolor{black}{cross Kerr interaction} and rabi frequency.
    The ground state and first excited state are 
    $|E_6\rangle $ and $|E_5\rangle $, respectively.
    $|E_1\rangle $, $|E_2\rangle $,
    $|E_3\rangle $,
    and $|E_4\rangle $ are the other excited states.
    Note that $|E_2\rangle $
    and $|E_3\rangle $ are nearly degenerate because of the condition of the large four-body interaction.
    }
    \label{diagram-four}
\end{figure}
We show our analytical results in this section.
The Hamiltonian in Eq.~\eqref{simpleh} is divided into the time-independent Hamiltonian $H_0$ and time-dependent Hamiltonian $H_{\rm{drive}}(t)$. We assume that the driving strength $\lambda_2$ is much smaller than the energy gap
between the initial state and target state to be excited.
In this case, we can treat $H_{\rm{drive}}(t)$ as a perturbation.
Let us assume that the initial state is 
the ground state
of $H_0$.
We can induce a transition from the initial state to the other energy eigenstates of $H_0$ if the following two conditions are satisfied \cite{sakurai2020modern}. First,
the driving frequency of $H_{\rm{drive}}(t)$ should be resonant with the energy difference between the ground state and another energy eigenstate.
Second, we should have a non-zero transition matrix such as $\langle e|H_{\rm{drive}}(t)|g\rangle \neq 0$ where $|g\rangle $ is the ground state and 
$|e\rangle $ is one of the other energy eigenstates of $H_0$.
Importantly, this approach allows us to investigate the eigenenergy of $H_0$, so we can study the energy structure of the system.

Furthermore, let us discuss where \textcolor{black}{the initial state is} a superposition between the 
\YM{eigenstates} of $H_0$. Primarily, we consider that the initial state is given as \YM{$|\psi_0\rangle =\alpha|E_n\rangle + \beta |E_m\rangle
$}, where $|E_n\rangle$ ($|E_m\rangle$) is the energy eigenstate with an eigenvalue of $E_n$ ($E_m$).
If 
the driving frequency of $H_{\rm{drive}}(t)$ is resonant with $E_n-E_j$
($E_m-E_j$)
and $\langle E_n|H_{\rm{drive}}(t)|E_j\rangle \neq 0$
($\langle E_m|H_{\rm{drive}}(t)|E_j\rangle \neq 0$)
is satisfied, a transition from 
\YM{$|E_n\rangle $ ($|E_m\rangle $)} to \YM{$|E_j\rangle$}
occurs \YM{where $|E_j\rangle $ is \textcolor{black}{another eigenstate} of $H_0$ with an eigenvalue of $E_j$}.

We can diagonalize the Hamiltonian $H_0$
as follows where the Hamiltonian can be block-diagonalized.
First, we consider a subspace spanned by $|\!\!\uparrow \uparrow \downarrow \downarrow \rangle $, $|\!\downarrow \uparrow \downarrow \downarrow \rangle $, $|\!\downarrow \downarrow \uparrow \uparrow \rangle $, and $|\!\uparrow \downarrow \uparrow \uparrow \rangle $.
Here, 
the four-body interaction induces transitions only between 
$|\!\!\uparrow \uparrow \downarrow \downarrow \rangle $ and $|\!\!\downarrow \downarrow \uparrow \uparrow \rangle $.
Also, there is a finite transition matrix element between  $|\!\!\uparrow \uparrow \downarrow \downarrow \rangle $ ($|\!\!\downarrow \downarrow \uparrow \uparrow \rangle$) and $|\!\!\downarrow \uparrow \downarrow \downarrow \rangle $ ($|\!\!\uparrow \downarrow \uparrow \uparrow \rangle$) due to the driving \YM{on the first quibt}. 
\textcolor{black}{Concerning} $|\!\!\uparrow \uparrow \downarrow \downarrow \rangle $, $|\!\!\downarrow \uparrow \downarrow \downarrow \rangle $, $|\!\!\downarrow \downarrow \uparrow \uparrow \rangle $, and $|\!\!\uparrow \downarrow \uparrow \uparrow \rangle $, there are no transitions
except these, and so we can consider this subspace.
\YM{It is worth mentioning that,
since we have $\omega_1+\omega_2=\omega_3+\omega_4$ and 
$\omega^{(1)}_{\rm{rot}}+\omega^{(2)}_{\rm{rot}}=\omega^{(3)}_{\rm{rot}}+\omega^{(4)}_{\rm{rot}}$, we obtain $\sum _{j=3}^4\frac{\omega_j-\omega^{(j)}_{\rm{rot}}}{2}=\frac{\omega_2-\omega^{(2)}_{\rm{rot}}}{2}$. By using this,}
we can simplify
the Hamiltonian in this subspace 
as
\begin{eqnarray}
\tilde{H}_0
&=&g(|\!\uparrow \uparrow \downarrow \downarrow \rangle \langle \downarrow \downarrow \uparrow \uparrow\!| + |\!\downarrow \downarrow \uparrow \uparrow \rangle \langle \uparrow \uparrow \downarrow \downarrow\!|)\nonumber \\
&+&\frac{\lambda}{2}(
|\!\downarrow \uparrow \downarrow \downarrow \rangle \langle \uparrow \uparrow \downarrow \downarrow\!|
+
|\!\uparrow \uparrow \downarrow \downarrow \rangle \langle \downarrow \uparrow \downarrow \downarrow\!|
)\nonumber \\
&+&\frac{\lambda}{2}(|\!\uparrow \downarrow \uparrow \uparrow \rangle
\langle \downarrow \downarrow \uparrow \uparrow \!|
+ |\!\downarrow \downarrow \uparrow \uparrow \rangle \langle \uparrow \downarrow \uparrow \uparrow\!|)\nonumber \\
&+&
J_+(|\!\uparrow \uparrow \downarrow \downarrow \rangle \langle \!\uparrow \uparrow \downarrow \downarrow\!|
+|\! \downarrow \downarrow \uparrow \uparrow\rangle \langle \! \downarrow \downarrow \uparrow \uparrow\!|)
\nonumber \\
&+&J_-(|\!\downarrow \uparrow \downarrow \downarrow \rangle \langle \!\downarrow \uparrow \downarrow \downarrow\!|
+|\! \uparrow \downarrow \uparrow \uparrow\rangle \langle \! \uparrow \downarrow \uparrow \uparrow\!|)
\end{eqnarray}
where $J_+=(J_{12}-J_{13}-J_{14}-J_{23}-J_{24}+J_{34})/4$
and $J_-=(-J_{12}+J_{13}+J_{14}-J_{23}-J_{24}+J_{34})/4$.

We consider diagonalizing this Hamiltonian such that $\Tilde{H}_0=\sum _{n=1}^4E_n |E_n\rangle \langle E_n|$.
We set $\tilde{J}_+=J_+ + J_-$ and $\tilde{J}_-=J_+ - J_-$.
We can represent the eigenstates as follows:
\begin{eqnarray}
    |E_1\rangle &=&C_1(|\!\uparrow \uparrow \downarrow \downarrow \rangle
     -\frac{g+\Tilde{J}_--\sqrt{(g+\Tilde{J}_-)^2 +\lambda ^2}}{\lambda} |\!\uparrow \downarrow \uparrow \uparrow \rangle \nonumber \\
      &+&\frac{\lambda}{g+\Tilde{J}_-+\sqrt{(g+\Tilde{J}_-)^2 +\lambda ^2}} |\!\downarrow \uparrow \downarrow \downarrow \rangle
    + |\!\downarrow \downarrow \uparrow \uparrow \rangle),
    \nonumber
     \\
            |E_2\rangle &=&C_2(-|\!\uparrow \uparrow \downarrow \downarrow \rangle
    +\frac{g-\Tilde{J}_-+ \sqrt{(g-\Tilde{J}_-)^2 +\lambda ^2}}{\lambda} |\!\uparrow \downarrow \uparrow \uparrow \rangle\nonumber \\
     &-& \frac{\lambda}{-g+\Tilde{J}_-+ \sqrt{(g-\Tilde{J}_-)^2 +\lambda ^2}}|\!\downarrow \uparrow \downarrow \downarrow \rangle
    + |\!\downarrow \downarrow \uparrow \uparrow \rangle),
     \nonumber \\
                |E_3\rangle &=&C_3(|\!\uparrow \uparrow \downarrow \downarrow \rangle
    -\frac{g+\Tilde{J}_-+ \sqrt{(g+\Tilde{J}_-)^2 +\lambda ^2}}{\lambda} |\!\uparrow \downarrow \uparrow \uparrow \rangle\nonumber \\
     &+& \frac{\lambda}{g+\Tilde{J}_-- \sqrt{(g+\Tilde{J}_-)^2 +\lambda ^2}}|\!\downarrow \uparrow \downarrow \downarrow \rangle
    + |\!\downarrow \downarrow \uparrow \uparrow \rangle),
     \nonumber \\
            |E_4\rangle &=&C_4(-
     |\!\uparrow \uparrow \downarrow \downarrow \rangle 
    -\frac{-g+\Tilde{J}_- + \sqrt{(g-\Tilde{J}_-)^2+\lambda^2}}{\lambda}|\!\uparrow \downarrow \uparrow \uparrow \rangle
    \nonumber \\
     &+& \frac{\lambda}{g-\Tilde{J}_-+\sqrt{(g-\Tilde{J}_-)^2 +\lambda^2}}|\!\downarrow \uparrow \downarrow \downarrow \rangle
    +|\!\downarrow \downarrow \uparrow \uparrow \rangle). \nonumber
\end{eqnarray}
where $C_i$ denotes a renormalization factor of the state $|E_i\rangle $ for $i=1,2,3,4$.  
The corresponding eigenvalues are as follows:
\begin{eqnarray}
    E_1&=&\frac{g+\Tilde{J}_++\sqrt{(g+\Tilde{J}_-)^2+\lambda^2}}{2},  \\
    E_2&=&\frac{-g+\Tilde{J}_++\sqrt{(g-\Tilde{J}_-)^2+\lambda^2}}{2},  \\
        E_3&=&\frac{g+\Tilde{J}_+-\sqrt{(g+\Tilde{J}_-)^2+\lambda^2}}{2},  \\
        E_4&=&\frac{-g+\Tilde{J}_+-\sqrt{(g-\Tilde{J}_-)^2+\lambda^2}}{2}. 
\end{eqnarray}
Let us assume that $\lambda$ is much smaller than the other parameters, and we can simplify the eigenstates and eigenenergies as follows: (see Appendix \ref{relabel} for the details.)
\begin{eqnarray}
        |E_1\rangle 
     &\simeq&
     C_1\Big{(}|\!\uparrow \uparrow \downarrow \downarrow \rangle
     +\frac{\lambda}{2(g+\Tilde{J}_-)} |\!\uparrow \downarrow \uparrow \uparrow \rangle \nonumber \\
      &+&\frac{\lambda}{2(g+\Tilde{J}_-)} |\!\downarrow \uparrow \downarrow \downarrow \rangle
    + |\!\downarrow \downarrow \uparrow \uparrow \rangle\Big{)},
\end{eqnarray}
\begin{eqnarray}
         |E_2\rangle 
     &\simeq&C_2\Big{(}-|\!\uparrow \uparrow \downarrow \downarrow \rangle
    +\frac{2(g-\Tilde{J}_-)}{\lambda} |\!\uparrow \downarrow \uparrow \uparrow \rangle\nonumber \\
     &-&\frac{2(g-\Tilde{J}_-)}{\lambda}|\!\downarrow \uparrow \downarrow \downarrow \rangle
    + |\!\downarrow \downarrow \uparrow \uparrow \rangle\Big{)}, 
\end{eqnarray}
\begin{eqnarray}
    |E_3\rangle 
     &\simeq &C_3\Big{(}|\!\uparrow \uparrow \downarrow \downarrow \rangle - \frac{2(g+\Tilde{J}_-)}{\lambda}|\!\uparrow \downarrow \uparrow \uparrow \rangle\nonumber \\
     &-&\frac{2(g+\Tilde{J}_-)}{\lambda}|\!\downarrow \uparrow \downarrow \downarrow \rangle+|\!\downarrow \downarrow \uparrow \uparrow \rangle\Big{)},
\end{eqnarray}
\begin{eqnarray}
    |E_4\rangle &\simeq &
    C_4\Big{(}-
     |\!\uparrow \uparrow \downarrow \downarrow \rangle 
    -\frac{\lambda}{2(g-\Tilde{J}_-)}|\!\uparrow \downarrow \uparrow \uparrow \rangle
    \nonumber \\
    &+&\frac{\lambda}{2(g-\Tilde{J}_-)}|\!\downarrow \uparrow \downarrow \downarrow \rangle
    +|\!\downarrow \downarrow \uparrow \uparrow \rangle\Big{)}.
\end{eqnarray}
\YM{
\textcolor{black}{The eigenstates are
$\frac{1}{\sqrt{2}}(|\!\uparrow \downarrow \uparrow \uparrow \rangle \pm |\!\downarrow \uparrow \downarrow \downarrow \rangle)$ or $\frac{1}{\sqrt{2}}(|\!\uparrow \uparrow \downarrow \downarrow \rangle \pm |\!\downarrow \downarrow \uparrow \uparrow \rangle)$ 
} \textcolor{black}{for small $\lambda$}. These are the dressed states formed by the four-body interaction.}

Second, we consider a subspace spanned by
$|\!\uparrow \downarrow \downarrow \downarrow \rangle $ and $|\!\downarrow\downarrow \downarrow \downarrow \rangle $ \textcolor{black}{where the Hamiltonian can be block-diadonalized}.
The Hamiltonian in this subspace is given by
\YM{
\begin{eqnarray}
 \tilde{H}'=(-\frac{\omega_2+\omega_3+\omega_4
 -\omega^{(2)}_{\rm{rot}}
 -\omega^{(3)}_{\rm{rot}}
-\omega^{(4)}_{\rm{rot}}
 }
 {2}+J_+')|\!\uparrow \downarrow \downarrow \downarrow \rangle \langle \uparrow \downarrow \downarrow \downarrow\!|\nonumber \\
 +(-\frac{\omega_2+\omega_3+\omega_4-\omega^{(2)}_{\rm{rot}}
 -\omega^{(3)}_{\rm{rot}}
-\omega^{(4)}_{\rm{rot}}}{2}+J_-')|\!\downarrow \downarrow \downarrow \downarrow \rangle \langle \downarrow \downarrow \downarrow \downarrow\!| \nonumber \\
+\frac{\lambda}{2} (|\!\uparrow \downarrow \downarrow \downarrow \rangle 
 \langle \downarrow \downarrow \downarrow \downarrow\!|
 +|\!\downarrow \downarrow \downarrow \downarrow \rangle \langle \uparrow \downarrow \downarrow \downarrow\!|), \nonumber
\end{eqnarray}}
where $J_+'=(-J_{12}-J_{13}-J_{14}+J_{23}+J_{24}+J_{34})/4$
and $J_-'=(J_{12}+J_{13}+J_{14}+J_{23}+J_{24}+J_{34})/4$.
By defining $\hat{\sigma}_z$ and $\hat{\sigma}_x$ as Pauli matrices in the subspace spanned by $|\!\uparrow \downarrow \downarrow \downarrow \rangle $ and $|\!\downarrow\downarrow \downarrow \downarrow \rangle $,
we can rewrite this Hamiltonian as
$\tilde{H}'=\frac{\epsilon}{2}\hat{\sigma}_z + \frac{\lambda}{2}\hat{\sigma}_x + \frac{k}{2}\hat{1}$ where $\epsilon=J_+'-J_-'$, 
and \YM{$k=-(\omega_2+\omega_3+\omega_4-\omega^{(2)}_{\rm{rot}}
 -\omega^{(3)}_{\rm{rot}}
-\omega^{(4)}_{\rm{rot}})+J_+'+J_-'$.}
The eigenstates are as follows:
\begin{eqnarray}
  &&  |E_5\rangle = \cos \theta |\!\uparrow \downarrow \downarrow \downarrow \rangle + \sin \theta |\!\downarrow \downarrow \downarrow \downarrow \rangle , \\
&&    |E_6\rangle = -\sin \theta |\!\uparrow \downarrow \downarrow \downarrow \rangle + \cos \theta |\! \downarrow \downarrow \downarrow \downarrow \rangle, 
\end{eqnarray}
where $r = \sqrt{\epsilon^2 + \lambda ^2}$, $\epsilon= r \cos 2\theta$, $\lambda= r\sin 2\theta$.
The corresponding eigenenergies are $E_5=(k+r)/2$ and $E_6=(k-r)/2$.

We consider possible transitions between the ground state and the excited states.
\textcolor{black}{We assume that the system relaxes into $|\! \!\downarrow \downarrow \downarrow \downarrow \rangle$ when there is no drive,} \YM{and we consider this as the initial state.
Since this state is written as}
$|\! \!\downarrow \downarrow \downarrow \downarrow \rangle =\sin \theta |E_5\rangle +\cos \theta |E_6\rangle $, we consider a transition matrix of 
$\langle E_n|H_{\rm{drive}}(t)|E_5\rangle $
and
$\langle E_n|H_{\rm{drive}}(t)|E_6\rangle $
for $n=1,2,3,4$.
\YM{The resonant conditions are $\omega _2'-\omega_{\rm{rot}}^{(2)}=E_n-E_5$ or $\omega _2'-\omega_{\rm{rot}}^{(2)}=E_n-E_6$ for $n=1,2,3,4$, and these do not depend on  $\omega _{\rm{rot}}^{(2)}$, $\omega _{\rm{rot}}^{(3)}$, or $\omega _{\rm{rot}}^{(4)}$ due to the conditions of $\omega _1+\omega _2=\omega _3+\omega _4$ and $\omega ^{(1)}_{\rm{rot}}+\omega ^{(2)}_{\rm{rot}}=\omega ^{(3)}_{\rm{rot}}+\omega ^{(4)}_{\rm{rot}}$.}
If this transition matrix is not zero, the corresponding transitions should occur.
This means that the initial state can be excited to $|E_1\rangle $, $|E_2\rangle $, $|E_3\rangle $, and $|E_4\rangle $.
We illustrate these transitions in Fig. \ref{diagram-four} where we assume $g\gg \lambda$.
Although there are
other excited states 
within the other subspaces,
we can ignore them
because the transition 
matrix elements from the initial state to such excited states become zero.

\section{Numerical results}

We show our numerical results.
By using the Hamiltonian in Eq.~\eqref{simpleh}, we solve the Lindblad master equation
from $t=0$ to $t=T$
with Lindblad operators of 
$\hat{L}_j=\sqrt{\gamma} \hat{\sigma}_- ^{(j)}$ $(j=1,2,3,4)$ as follows:
\begin{eqnarray}
    &&\frac{d\rho(t)}{dt}=-i[H,\rho]\nonumber \\
    &+&\sum_{j=1}^4 \frac{1}{2}
    (2\hat{L}_j \rho(t) \hat{L}^{\dagger}_j-\rho(t) \hat{L}_j^{\dagger}\hat{L}_j
    -\hat{L}_j^{\dagger}\hat{L}_j\rho (t)
    ), 
\end{eqnarray}
where $\rho (t)$ denotes a density matrix of the system and $\gamma$ denotes the decoherence rate.
The initial state is $|\! \!\downarrow \downarrow \downarrow \downarrow \rangle$.
\YM{We set $\omega_{\rm{rot}}^{(j)}=\omega _j$ for all $j$, and}
we plot an excitation probability
of the fourth qubit described as
$P_{\rm{e}}=\rm{Tr}[\rho(T)(\hat{\openone} +\hat{\sigma}_z^{(4)})/2]$
against the detuning \YM{$\Delta=\omega _2'-\omega _2$.} 
Importantly, since we do not drive the fourth qubit by microwave fields, the fourth qubit remains in the ground state without four-body interaction.

\begin{figure}[h!]
    \centering
    \includegraphics[width = 9.2cm]{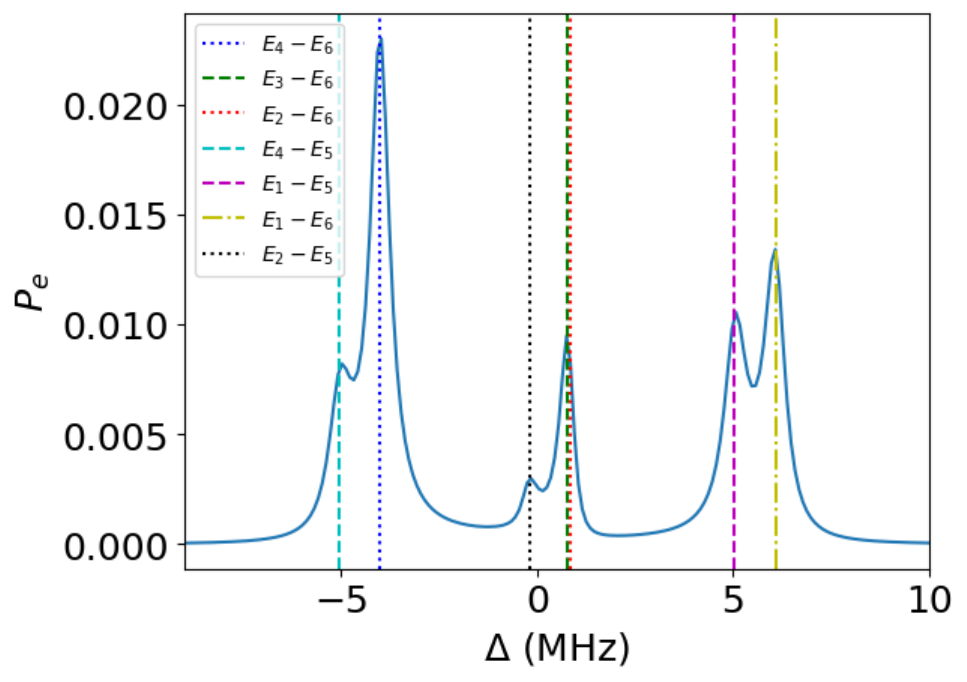}
    \caption{We plot 
     an excitation probability
of the fourth qubit described as
    $P_{\rm{e}}$
    where $x$ axis denotes the detuning \YM{$\Delta=\omega _2'-\omega _2$}.
     \YM{
    We also plot vertical lines as the difference between the energy eigenvalues of the Hamiltonian.}
    We set 
\YM{$\omega_1=\omega_1'=5.0$ GHz,}
    \YM{$\omega _2-\omega_1=13$ MHz, $\omega _3-\omega_1=-17$ MHz,
    $\omega _4-\omega_1=30$ MHz,} 
    $g=5.0$ MHz, $\lambda=1.0$ MHz, $\lambda_2=0.01$ MHz, $\gamma=0.25$ MHz, $J_{12}=-0.10$ MHz, $J_{13}=-0.42$ MHz, $J_{14}=-0.16$ MHz, $J_{23}=-0.40$ MHz, $J_{24}=-0.40$ MHz, $J_{34}=-0.61$ MHz,  and 
    $T=30 \ \mu s$. 
    }
    \label{newsteady-four-twobody-weak}
\end{figure}

First, let us consider a simple case that the four-body interaction is much stronger than the two-body interaction.
Here, to suppress the effect of the power broadening, we set $\lambda_2 < \gamma$.
We observe six peaks as shown in Fig.~\ref{newsteady-four-twobody-weak}.
The detuning \textcolor{black}{corresponding to} the peaks are 
\YM{$-5$ MHz, $-4$ MHz, $0$ MHz, $1$ MHz, \textcolor{black}{5 MHz}, and $6$ MHz.}
From the analytical solution,
we calculate
\YM{$E_4-E_5\simeq -5$ MHz, $E_4-E_6\simeq -4$ MHz, $E_3-E_5 \simeq 0$ MHz,
$E_2-E_5 \simeq 0$ MHz, $E_3-E_6 \simeq 14$ MHz, $E_2-E_6 \simeq 1$ MHz,
$E_1-E_5 \simeq 5$ MHz, and
$E_1-E_6 \simeq 6$ MHz.}
Since the transition energy of $E_3-E_5$
($E_3-E_6$)
is close to that of $E_2-E_5$ ($E_2-E_6$), we cannot resolve these two in the numerical simulations. 

\begin{figure}[h!]
    \centering
    \includegraphics[width = 9.2cm]{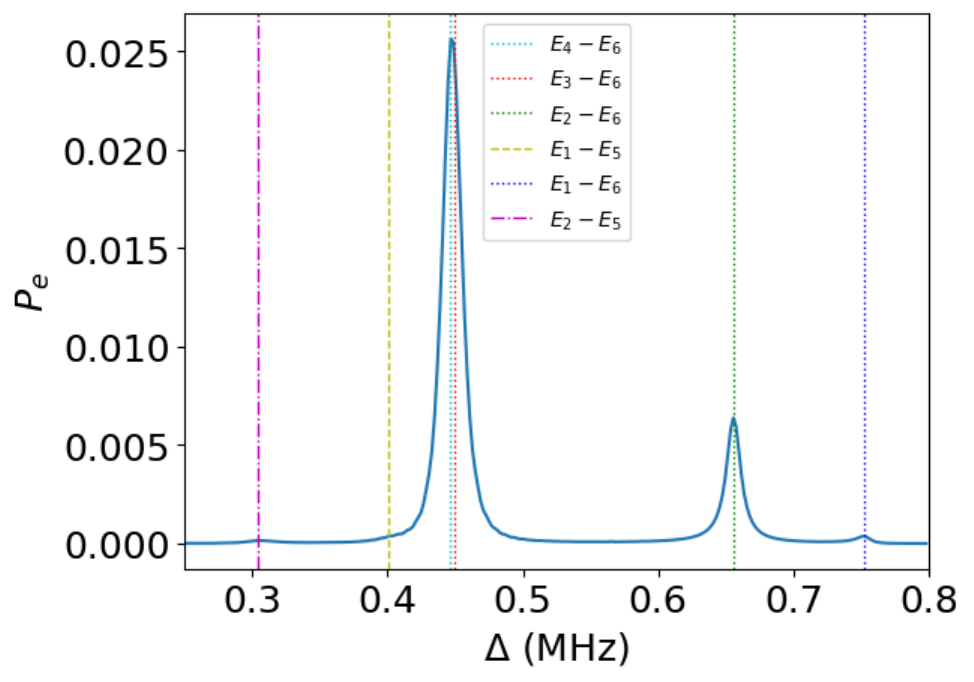}
    \caption{We plot an excitation probability
of the fourth qubit described as $P_{\rm{e}}$
    where $x$ axis denotes the detuning $\Delta$.
    \YM{
    We also plot vertical lines
    as the difference between the energy eigenvalues of the Hamiltonian.}
     We set 
\YM{$\omega_1=\omega_1'=5.0$ GHz,}
    \YM{$\omega _2-\omega_1=13$ MHz, $\omega _3-\omega_1=-17$ MHz,
    $\omega _4-\omega_1=30$ MHz,} 
    \AdMORI{$g=0.05$} MHz, 
    \AdMORI{$\lambda = 0.090$} MHz, 
    \AdMORI{$\lambda_2=0.0075$} MHz, 
    \AdMORI{$\gamma = 0.0060$} MHz, 
    \AdMORI{$J_{12}=-0.10$} MHz, 
    \AdMORI{$J_{13}=-0.42$} MHz, 
    \AdMORI{$J_{14}=-0.16$} MHz, 
    \AdMORI{$J_{23}=-0.40$} MHz, 
    \AdMORI{ $J_{24}=-0.40$} MHz, 
    \AdMORI{$J_{34}=-0.61$} MHz,  and 
    \AdMORI{$T=1000 \ \mu s$}.
    }
    \label{3steady-four-twobody-weak}
\end{figure}


Next, let us consider 
\YM{a more realistic} case that the strength of the four-body interaction is comparable with that of the two-body interaction \cite{Puri2017_npjq}.
\YM{Also, $g$, $\lambda$, and $\gamma$ are set to be smaller than
those
in Fig.~\ref{newsteady-four-twobody-weak}.}
\textcolor{black}{
We observe two main peak as shown in Fig.~\ref{3steady-four-twobody-weak}.
The detuning corresponding to the peaks are 
0.45 MHz and 0.66 MHz.
}
From the analytical solution, we calculate 
\textcolor{black}{$E_3-E_6 \simeq 0.45$ MHz,  $E_4-E_6 \simeq 0.45$ MHz, and $E_2-E_6 \simeq 0.66$ MHz.}
Here, the transition energy of $E_3-E_6$
is close to that of $E_4-E_6$, and we cannot resolve these two in the numerical simulations. 
In this parameter regime,
the initial state is approximately given as $|\! \!\downarrow \downarrow \downarrow \downarrow \rangle \simeq |E_6\rangle $.
     This means that we can ignore the transition 
     from $|E_5\rangle $, 
     \textcolor{black}{which is consistent with the fact that we cannot observe the transitions from $|E_5\rangle $ in the numerical results.}

\section{Conclusion}
In conclusion, we study the spectroscopic measurements of the Kerr non-linear resonators with four-body interaction.
We strongly drive one of the resonators while weakly driving another. 
When we measure the population of the excited state of a resonator that we do not drive, we observe peaks at six frequencies. By analyzing the peak frequencies, we can study the energy structure of the system.
 Our spectroscopy method does not require fast pulsed operations, and 
 our method can be performed with continuous-wave measurements.
Our method helps characterize a device if the four-body coupling strength is realized, which is crucial for quantum annealing with KPOs.

\begin{acknowledgments}
We would like to thank Takaaki Aoki and Ryoji Miyazaki for the fruitful comments. 
This paper is based
on the results obtained from a project, JPNP16007, commissioned by the New Energy and
Industrial Technology Development Organization (NEDO). This work was partially supported
by JST Moonshot R\&D (Grant Number JPMJMS226C).
Y. Matsuzaki is supported by JSPS KAKENHI (Grant Number
23H04390). This work was also supported by CREST
(JPMJCR23I5), JST.

\end{acknowledgments}

\appendix
\section{Analytical form of the energy eigenstates.}\label{relabel}
We have
\begin{eqnarray}
    |E_1\rangle &=&C_1(|\!\uparrow \uparrow \downarrow \downarrow \rangle
     -\frac{g+\Tilde{J}_--\sqrt{(g+\Tilde{J}_-)^2 +\lambda ^2}}{\lambda} |\!\uparrow \downarrow \uparrow \uparrow \rangle \nonumber \\
      &+&\frac{\lambda}{g+\Tilde{J}_-+\sqrt{(g+\Tilde{J}_-)^2 +\lambda ^2}} |\!\downarrow \uparrow \downarrow \downarrow \rangle
    + |\!\downarrow \downarrow \uparrow \uparrow \rangle),
    \nonumber
     \\
            |E_2\rangle &=&C_2(-|\!\uparrow \uparrow \downarrow \downarrow \rangle
    +\frac{g-\Tilde{J}_-+ \sqrt{(g-\Tilde{J}_-)^2 +\lambda ^2}}{\lambda} |\!\uparrow \downarrow \uparrow \uparrow \rangle\nonumber \\
     &-& \frac{\lambda}{-g+\Tilde{J}_-+ \sqrt{(g-\Tilde{J}_-)^2 +\lambda ^2}}|\!\downarrow \uparrow \downarrow \downarrow \rangle
    + |\!\downarrow \downarrow \uparrow \uparrow \rangle),
     \nonumber \\
                |E_3\rangle &=&C_3(|\!\uparrow \uparrow \downarrow \downarrow \rangle
    -\frac{g+\Tilde{J}_-+ \sqrt{(g+\Tilde{J}_-)^2 +\lambda ^2}}{\lambda} |\!\uparrow \downarrow \uparrow \uparrow \rangle\nonumber \\
     &+& \frac{\lambda}{g+\Tilde{J}_-- \sqrt{(g+\Tilde{J}_-)^2 +\lambda ^2}}|\!\downarrow \uparrow \downarrow \downarrow \rangle
    + |\!\downarrow \downarrow \uparrow \uparrow \rangle),
     \nonumber \\
            |E_4\rangle &=&C_4(-
     |\!\uparrow \uparrow \downarrow \downarrow \rangle 
    -\frac{-g+\Tilde{J}_- + \sqrt{(g-\Tilde{J}_-)^2+\lambda^2}}{\lambda}|\!\uparrow \downarrow \uparrow \uparrow \rangle
    \nonumber \\
     &+& \frac{\lambda}{g-\Tilde{J}_-+\sqrt{(g-\Tilde{J}_-)^2 +\lambda^2}}|\!\downarrow \uparrow \downarrow \downarrow \rangle
    +|\!\downarrow \downarrow \uparrow \uparrow \rangle). \nonumber
\end{eqnarray}
The corresponding eigenvalues are as follows:
\begin{eqnarray}
E_1&=&\frac{g+\Tilde{J}_++\sqrt{(g+\Tilde{J}_-)^2+\lambda^2}}{2},  \\
    E_2&=&\frac{-g+\Tilde{J}_++\sqrt{(g-\Tilde{J}_-)^2+\lambda^2}}{2},  \\
        E_3&=&\frac{g+\Tilde{J}_+-\sqrt{(g+\Tilde{J}_-)^2+\lambda^2}}{2},  \\
        E_4&=&\frac{-g+\Tilde{J}_+-\sqrt{(g-\Tilde{J}_-)^2+\lambda^2}}{2}. 
\end{eqnarray}
We consider a case with $\lambda \ll g$, and will derive simplified forms of the energy eigenstates.
\subsection{$|E_2\rangle $ and $|E_4\rangle $}
Let us consider $|E_2\rangle $ and $|E_4\rangle $.
We have $\sqrt{(g-\Tilde{J}_-)^2 +\lambda ^2}\simeq 
|g-\Tilde{J}_-|(1+\frac{\lambda^2}{2|g-\Tilde{J}_-|^2})=|g-\Tilde{J}_-|+\frac{\lambda^2}{2|g-\Tilde{J}_-|}$. First, we assume $g-\Tilde{J}_-\geq 0$. In this case, we obtain 
\begin{eqnarray}
     |E_2\rangle &=&C_2(-|\!\uparrow \uparrow \downarrow \downarrow \rangle
    +\frac{g-\Tilde{J}_-+ \sqrt{(g-\Tilde{J}_-)^2 +\lambda ^2}}{\lambda} |\!\uparrow \downarrow \uparrow \uparrow \rangle\nonumber \\
     &-& \frac{\lambda}{-g+\Tilde{J}_-+ \sqrt{(g-\Tilde{J}_-)^2 +\lambda ^2}}|\!\downarrow \uparrow \downarrow \downarrow \rangle
    + |\!\downarrow \downarrow \uparrow \uparrow \rangle),
     \nonumber \\
     &\simeq&C_2(-|\!\uparrow \uparrow \downarrow \downarrow \rangle
    +\frac{2(g-\Tilde{J}_-)}{\lambda} |\!\uparrow \downarrow \uparrow \uparrow \rangle\nonumber \\
     &-&\frac{2(g-\Tilde{J}_-)}{\lambda}|\!\downarrow \uparrow \downarrow \downarrow \rangle
    + |\!\downarrow \downarrow \uparrow \uparrow \rangle), \nonumber \\
    |E_4\rangle &=&C_4(-
     |\!\uparrow \uparrow \downarrow \downarrow \rangle 
    -\frac{-g+\Tilde{J}_- + \sqrt{(g-\Tilde{J}_-)^2+\lambda^2}}{\lambda}|\!\uparrow \downarrow \uparrow \uparrow \rangle
    \nonumber \\
     &+& \frac{\lambda}{g-\Tilde{J}_-+\sqrt{(g-\Tilde{J}_-)^2 +\lambda^2}}|\!\downarrow \uparrow \downarrow \downarrow \rangle
    +|\!\downarrow \downarrow \uparrow \uparrow \rangle) \nonumber \\
    &\simeq &
    C_4(-
     |\!\uparrow \uparrow \downarrow \downarrow \rangle 
    -\frac{\lambda}{2(g-\Tilde{J}_-)}|\!\uparrow \downarrow \uparrow \uparrow \rangle
    \nonumber \\
    &+&\frac{\lambda}{2(g-\Tilde{J}_-)}|\!\downarrow \uparrow \downarrow \downarrow \rangle
    +|\!\downarrow \downarrow \uparrow \uparrow \rangle)
\end{eqnarray}
The corresponding eigenenergies are $E_2=\frac{\Tilde{J}_+ -\Tilde{J}_-}{2}$ and $E_4=-g+\frac{\Tilde{J}_+ + \Tilde{J}_-}{2}$.
Second, we assume $g-\Tilde{J}_-< 0$. In this case, we obtain 
\begin{eqnarray}
     |E_2\rangle &=&C_2(-|\!\uparrow \uparrow \downarrow \downarrow \rangle
    +\frac{g-\Tilde{J}_-+ \sqrt{(g-\Tilde{J}_-)^2 +\lambda ^2}}{\lambda} |\!\uparrow \downarrow \uparrow \uparrow \rangle\nonumber \\
     &-& \frac{\lambda}{-g+\Tilde{J}_-+ \sqrt{(g-\Tilde{J}_-)^2 +\lambda ^2}}|\!\downarrow \uparrow \downarrow \downarrow \rangle
    + |\!\downarrow \downarrow \uparrow \uparrow \rangle),
     \nonumber \\
     &\simeq&C_2(-|\!\uparrow \uparrow \downarrow \downarrow \rangle
    -\frac{\lambda}{2(g-\Tilde{J}_-)} |\!\uparrow \downarrow \uparrow \uparrow \rangle\nonumber \\
     &+&\frac{\lambda}{2(g-\Tilde{J}_-)}|\!\downarrow \uparrow \downarrow \downarrow \rangle
    + |\!\downarrow \downarrow \uparrow \uparrow \rangle), \nonumber \\
    |E_4\rangle &=&C_4(-
     |\!\uparrow \uparrow \downarrow \downarrow \rangle 
    -\frac{-g+\Tilde{J}_- + \sqrt{(g-\Tilde{J}_-)^2+\lambda^2}}{\lambda}|\!\uparrow \downarrow \uparrow \uparrow \rangle
    \nonumber \\
     &+& \frac{\lambda}{g-\Tilde{J}_-+\sqrt{(g-\Tilde{J}_-)^2 +\lambda^2}}|\!\downarrow \uparrow \downarrow \downarrow \rangle
    +|\!\downarrow \downarrow \uparrow \uparrow \rangle) \nonumber \\
    &\simeq &
    C_4(-
     |\!\uparrow \uparrow \downarrow \downarrow \rangle 
    +\frac{2(g-\Tilde{J}_-)}{\lambda}|\!\uparrow \downarrow \uparrow \uparrow \rangle
    \nonumber \\
    &-&\frac{2(g-\Tilde{J}_-)}{\lambda}|\!\downarrow \uparrow \downarrow \downarrow \rangle
    +|\!\downarrow \downarrow \uparrow \uparrow \rangle)
\end{eqnarray}
The corresponding eigenenergies  are 
$E_2=-g+\frac{\Tilde{J}_+ + \Tilde{J}_-}{2}$ and
$E_4=\frac{\Tilde{J}_+ - \Tilde{J}_-}{2}$.
So $|E_2\rangle $ ($E_2$)  for $g+\tilde{J}_-\geq 0$  has the same form as $|E_4\rangle $ $(E_4)$  for $g+\tilde{J}_-< 0$. Also, $|E_2\rangle $ $(E_2)$ for $g+\tilde{J}_-\geq 0$  has the same form as $|E_4\rangle $ $(E_4)$  for $g+\tilde{J}_-< 0$. 
To unify the notation,
we relabel $|E_2\rangle $ ($E_2$) with $|E_4\rangle $ ($E_4$) for $g+\tilde{J}_-< 0$.

\subsection{$|E_1\rangle $ and $|E_3\rangle $}
Let us consider $|E_1\rangle $ and $|E_3\rangle $.
We have $\sqrt{(g+\Tilde{J}_-)^2 +\lambda ^2}\simeq 
|g+\Tilde{J}_-|(1+\frac{\lambda^2}{2|g+\Tilde{J}_-|^2})=|g+\Tilde{J}_-|+\frac{\lambda^2}{2|g+\Tilde{J}_-|}$. First, we assume $g+\Tilde{J}_-\geq 0$. In this case, we obtain  
\begin{eqnarray}
        |E_1\rangle &=&C_1(|\!\uparrow \uparrow \downarrow \downarrow \rangle
     -\frac{g+\Tilde{J}_--\sqrt{(g+\Tilde{J}_-)^2 +\lambda ^2}}{\lambda} |\!\uparrow \downarrow \uparrow \uparrow \rangle \nonumber \\
      &+&\frac{\lambda}{g+\Tilde{J}_-+\sqrt{(g+\Tilde{J}_-)^2 +\lambda ^2}} |\!\downarrow \uparrow \downarrow \downarrow \rangle
    + |\!\downarrow \downarrow \uparrow \uparrow \rangle),
    \nonumber
     \\
     &\simeq&
     C_1(|\!\uparrow \uparrow \downarrow \downarrow \rangle
     +\frac{\lambda}{2(g+\Tilde{J}_-)} |\!\uparrow \downarrow \uparrow \uparrow \rangle \nonumber \\
      &+&\frac{\lambda}{2(g+\Tilde{J}_-)} |\!\downarrow \uparrow \downarrow \downarrow \rangle
    + |\!\downarrow \downarrow \uparrow \uparrow \rangle),
    \nonumber
\end{eqnarray}
\begin{eqnarray}
    |E_3\rangle &=&C_3(|\!\uparrow \uparrow \downarrow \downarrow \rangle
    -\frac{g+\Tilde{J}_-+ \sqrt{(g+\Tilde{J}_-)^2 +\lambda ^2}}{\lambda} |\!\uparrow \downarrow \uparrow \uparrow \rangle\nonumber \\
     &+& \frac{\lambda}{g+\Tilde{J}_-- \sqrt{(g+\Tilde{J}_-)^2 +\lambda ^2}}|\!\downarrow \uparrow \downarrow \downarrow \rangle
    + |\!\downarrow \downarrow \uparrow \uparrow \rangle),
     \nonumber \\
     &\simeq &C_3(|\!\uparrow \uparrow \downarrow \downarrow \rangle - \frac{2(g+\Tilde{J}_-)}{\lambda}|\!\uparrow \downarrow \uparrow \uparrow \rangle\nonumber \\
     &-&\frac{2(g+\Tilde{J}_-)}{\lambda}|\!\downarrow \uparrow \downarrow \downarrow \rangle+|\!\downarrow \downarrow \uparrow \uparrow \rangle)
\end{eqnarray}
where the corresponding eigenvalues are $E_1\simeq g+\frac{\Tilde{J}_+ + \Tilde{J}_-}{2}$ and $E_3\simeq \frac{\Tilde{J}_+ - \Tilde{J}_-}{2}$.
Second, we assume $g+\Tilde{J}_-< 0$. In this case, we obtain  
\begin{eqnarray}
        |E_1\rangle &=&C_1(|\!\uparrow \uparrow \downarrow \downarrow \rangle
     -\frac{g+\Tilde{J}_--\sqrt{(g+\Tilde{J}_-)^2 +\lambda ^2}}{\lambda} |\!\uparrow \downarrow \uparrow \uparrow \rangle \nonumber \\
      &+&\frac{\lambda}{g+\Tilde{J}_-+\sqrt{(g+\Tilde{J}_-)^2 +\lambda ^2}} |\!\downarrow \uparrow \downarrow \downarrow \rangle
    + |\!\downarrow \downarrow \uparrow \uparrow \rangle),
    \nonumber
     \\
     &\simeq&
     C_1(|\!\uparrow \uparrow \downarrow \downarrow \rangle
     -\frac{2(g+\Tilde{J}_-)}{\lambda} |\!\uparrow \downarrow \uparrow \uparrow \rangle \nonumber \\
      &-&\frac{2(g+\Tilde{J}_-)}{\lambda} |\!\downarrow \uparrow \downarrow \downarrow \rangle
    + |\!\downarrow \downarrow \uparrow \uparrow \rangle),
    \nonumber
\end{eqnarray}
\begin{eqnarray}
    |E_3\rangle &=&C_3(|\!\uparrow \uparrow \downarrow \downarrow \rangle
    -\frac{g+\Tilde{J}_-+ \sqrt{(g+\Tilde{J}_-)^2 +\lambda ^2}}{\lambda} |\!\uparrow \downarrow \uparrow \uparrow \rangle\nonumber \\
     &+& \frac{\lambda}{g+\Tilde{J}_-- \sqrt{(g+\Tilde{J}_-)^2 +\lambda ^2}}|\!\downarrow \uparrow \downarrow \downarrow \rangle
    + |\!\downarrow \downarrow \uparrow \uparrow \rangle),
     \nonumber \\
     &\simeq &C_3(|\!\uparrow \uparrow \downarrow \downarrow \rangle + \frac{\lambda}{2(g+\Tilde{J}_-)}|\!\uparrow \downarrow \uparrow \uparrow \rangle\nonumber \\
     &-&\frac{\lambda}{2(g+\Tilde{J}_-)}|\!\downarrow \uparrow \downarrow \downarrow \rangle+|\!\downarrow \downarrow \uparrow \uparrow \rangle)
\end{eqnarray}
where the corresponding eigenenergies are $E_1\simeq \frac{\Tilde{J}_+ -\Tilde{J}_-}{2}$ and $E_3=g+\frac{\Tilde{J}_+ +\Tilde{J}_-}{2}$.
So $|E_3\rangle $ ($E_3$)  for $g+\tilde{J}_-\geq 0$  has the same form as $|E_1\rangle $ $(E_1)$  for $g+\tilde{J}_-< 0$. Also, $|E_1\rangle $ $(E_1)$ for $g+\tilde{J}_-\geq 0$  has the same form as $|E_3\rangle $ $(E_3)$  for $g+\tilde{J}_-< 0$. 
To unify the notation,
we relabel $|E_1\rangle $ ($E_1$) with $|E_3\rangle $ ($E_3$) for $g+\tilde{J}_-< 0$.

\bibliography{ref}

\begin{thebibliography}{31}%
\makeatletter
\providecommand \@ifxundefined [1]{%
 \@ifx{#1\undefined}
}%
\providecommand \@ifnum [1]{%
 \ifnum #1\expandafter \@firstoftwo
 \else \expandafter \@secondoftwo
 \fi
}%
\providecommand \@ifx [1]{%
 \ifx #1\expandafter \@firstoftwo
 \else \expandafter \@secondoftwo
 \fi
}%
\providecommand \natexlab [1]{#1}%
\providecommand \enquote  [1]{``#1''}%
\providecommand \bibnamefont  [1]{#1}%
\providecommand \bibfnamefont [1]{#1}%
\providecommand \citenamefont [1]{#1}%
\providecommand \href@noop [0]{\@secondoftwo}%
\providecommand \href [0]{\begingroup \@sanitize@url \@href}%
\providecommand \@href[1]{\@@startlink{#1}\@@href}%
\providecommand \@@href[1]{\endgroup#1\@@endlink}%
\providecommand \@sanitize@url [0]{\catcode `\\12\catcode `\$12\catcode `\&12\catcode `\#12\catcode `\^12\catcode `\_12\catcode `\%12\relax}%
\providecommand \@@startlink[1]{}%
\providecommand \@@endlink[0]{}%
\providecommand \url  [0]{\begingroup\@sanitize@url \@url }%
\providecommand \@url [1]{\endgroup\@href {#1}{\urlprefix }}%
\providecommand \urlprefix  [0]{URL }%
\providecommand \Eprint [0]{\href }%
\providecommand \doibase [0]{https://doi.org/}%
\providecommand \selectlanguage [0]{\@gobble}%
\providecommand \bibinfo  [0]{\@secondoftwo}%
\providecommand \bibfield  [0]{\@secondoftwo}%
\providecommand \translation [1]{[#1]}%
\providecommand \BibitemOpen [0]{}%
\providecommand \bibitemStop [0]{}%
\providecommand \bibitemNoStop [0]{.\EOS\space}%
\providecommand \EOS [0]{\spacefactor3000\relax}%
\providecommand \BibitemShut  [1]{\csname bibitem#1\endcsname}%
\let\auto@bib@innerbib\@empty
\bibitem [{\citenamefont {Kadowaki}\ and\ \citenamefont {Nishimori}(1998)}]{Kadowaki_1998_pre}%
  \BibitemOpen
  \bibfield  {author} {\bibinfo {author} {\bibfnamefont {T.}~\bibnamefont {Kadowaki}}\ and\ \bibinfo {author} {\bibfnamefont {H.}~\bibnamefont {Nishimori}},\ }\bibfield  {title} {\bibinfo {title} {Quantum annealing in the transverse ising model},\ }\href@noop {} {\bibfield  {journal} {\bibinfo  {journal} {Physical Review E}\ }\textbf {\bibinfo {volume} {58}},\ \bibinfo {pages} {5355} (\bibinfo {year} {1998})}\BibitemShut {NoStop}%
\bibitem [{\citenamefont {Farhi}\ \emph {et~al.}(2000)\citenamefont {Farhi}, \citenamefont {Goldstone}, \citenamefont {Gutmann},\ and\ \citenamefont {Sipser}}]{farhi2000quantum}%
  \BibitemOpen
  \bibfield  {author} {\bibinfo {author} {\bibfnamefont {E.}~\bibnamefont {Farhi}}, \bibinfo {author} {\bibfnamefont {J.}~\bibnamefont {Goldstone}}, \bibinfo {author} {\bibfnamefont {S.}~\bibnamefont {Gutmann}},\ and\ \bibinfo {author} {\bibfnamefont {M.}~\bibnamefont {Sipser}},\ }\bibfield  {title} {\bibinfo {title} {Quantum computation by adiabatic evolution},\ }\href@noop {} {\bibfield  {journal} {\bibinfo  {journal} {arXiv preprint}\ } (\bibinfo {year} {2000})}\BibitemShut {NoStop}%
\bibitem [{\citenamefont {Farhi}\ \emph {et~al.}(2001)\citenamefont {Farhi}, \citenamefont {Goldstone}, \citenamefont {Gutmann}, \citenamefont {Lapan}, \citenamefont {Lundgren},\ and\ \citenamefont {Preda}}]{farhi2001quantum}%
  \BibitemOpen
  \bibfield  {author} {\bibinfo {author} {\bibfnamefont {E.}~\bibnamefont {Farhi}}, \bibinfo {author} {\bibfnamefont {J.}~\bibnamefont {Goldstone}}, \bibinfo {author} {\bibfnamefont {S.}~\bibnamefont {Gutmann}}, \bibinfo {author} {\bibfnamefont {J.}~\bibnamefont {Lapan}}, \bibinfo {author} {\bibfnamefont {A.}~\bibnamefont {Lundgren}},\ and\ \bibinfo {author} {\bibfnamefont {D.}~\bibnamefont {Preda}},\ }\bibfield  {title} {\bibinfo {title} {A quantum adiabatic evolution algorithm applied to random instances of an np-complete problem},\ }\href@noop {} {\bibfield  {journal} {\bibinfo  {journal} {Science}\ }\textbf {\bibinfo {volume} {292}},\ \bibinfo {pages} {472} (\bibinfo {year} {2001})}\BibitemShut {NoStop}%
\bibitem [{\citenamefont {Choi}(2011)}]{choi2011minor}%
  \BibitemOpen
  \bibfield  {author} {\bibinfo {author} {\bibfnamefont {V.}~\bibnamefont {Choi}},\ }\bibfield  {title} {\bibinfo {title} {Minor-embedding in adiabatic quantum computation: Ii. minor-universal graph design},\ }\href@noop {} {\bibfield  {journal} {\bibinfo  {journal} {Quantum Information Processing}\ }\textbf {\bibinfo {volume} {10}},\ \bibinfo {pages} {343} (\bibinfo {year} {2011})}\BibitemShut {NoStop}%
\bibitem [{\citenamefont {Childs}\ \emph {et~al.}(2001)\citenamefont {Childs}, \citenamefont {Farhi},\ and\ \citenamefont {Preskill}}]{childs2001robustness}%
  \BibitemOpen
  \bibfield  {author} {\bibinfo {author} {\bibfnamefont {A.~M.}\ \bibnamefont {Childs}}, \bibinfo {author} {\bibfnamefont {E.}~\bibnamefont {Farhi}},\ and\ \bibinfo {author} {\bibfnamefont {J.}~\bibnamefont {Preskill}},\ }\bibfield  {title} {\bibinfo {title} {Robustness of adiabatic quantum computation},\ }\href@noop {} {\bibfield  {journal} {\bibinfo  {journal} {Physical Review A}\ }\textbf {\bibinfo {volume} {65}},\ \bibinfo {pages} {012322} (\bibinfo {year} {2001})}\BibitemShut {NoStop}%
\bibitem [{\citenamefont {Morita}\ and\ \citenamefont {Nishimori}(2008)}]{morita2008mathematical}%
  \BibitemOpen
  \bibfield  {author} {\bibinfo {author} {\bibfnamefont {S.}~\bibnamefont {Morita}}\ and\ \bibinfo {author} {\bibfnamefont {H.}~\bibnamefont {Nishimori}},\ }\bibfield  {title} {\bibinfo {title} {Mathematical foundation of quantum annealing},\ }\href@noop {} {\bibfield  {journal} {\bibinfo  {journal} {Journal of Mathematical Physics}\ }\textbf {\bibinfo {volume} {49}} (\bibinfo {year} {2008})}\BibitemShut {NoStop}%
\bibitem [{\citenamefont {Harris}\ \emph {et~al.}(2010{\natexlab{a}})\citenamefont {Harris}, \citenamefont {Johansson}, \citenamefont {Berkley}, \citenamefont {Johnson}, \citenamefont {Lanting}, \citenamefont {Han}, \citenamefont {Bunyk}, \citenamefont {Ladizinsky}, \citenamefont {Oh}, \citenamefont {Perminov} \emph {et~al.}}]{harris2010experimental}%
  \BibitemOpen
  \bibfield  {author} {\bibinfo {author} {\bibfnamefont {R.}~\bibnamefont {Harris}}, \bibinfo {author} {\bibfnamefont {J.}~\bibnamefont {Johansson}}, \bibinfo {author} {\bibfnamefont {A.}~\bibnamefont {Berkley}}, \bibinfo {author} {\bibfnamefont {M.}~\bibnamefont {Johnson}}, \bibinfo {author} {\bibfnamefont {T.}~\bibnamefont {Lanting}}, \bibinfo {author} {\bibfnamefont {S.}~\bibnamefont {Han}}, \bibinfo {author} {\bibfnamefont {P.}~\bibnamefont {Bunyk}}, \bibinfo {author} {\bibfnamefont {E.}~\bibnamefont {Ladizinsky}}, \bibinfo {author} {\bibfnamefont {T.}~\bibnamefont {Oh}}, \bibinfo {author} {\bibfnamefont {I.}~\bibnamefont {Perminov}}, \emph {et~al.},\ }\bibfield  {title} {\bibinfo {title} {Experimental demonstration of a robust and scalable flux qubit},\ }\href@noop {} {\bibfield  {journal} {\bibinfo  {journal} {Physical Review B}\ }\textbf {\bibinfo {volume} {81}},\ \bibinfo {pages} {134510} (\bibinfo {year} {2010}{\natexlab{a}})}\BibitemShut {NoStop}%
\bibitem [{\citenamefont {Harris}\ \emph {et~al.}(2010{\natexlab{b}})\citenamefont {Harris}, \citenamefont {Johnson}, \citenamefont {Lanting}, \citenamefont {Berkley}, \citenamefont {Johansson}, \citenamefont {Bunyk}, \citenamefont {Tolkacheva}, \citenamefont {Ladizinsky}, \citenamefont {Ladizinsky}, \citenamefont {Oh} \emph {et~al.}}]{harris2010experimental2}%
  \BibitemOpen
  \bibfield  {author} {\bibinfo {author} {\bibfnamefont {R.}~\bibnamefont {Harris}}, \bibinfo {author} {\bibfnamefont {M.~W.}\ \bibnamefont {Johnson}}, \bibinfo {author} {\bibfnamefont {T.}~\bibnamefont {Lanting}}, \bibinfo {author} {\bibfnamefont {A.}~\bibnamefont {Berkley}}, \bibinfo {author} {\bibfnamefont {J.}~\bibnamefont {Johansson}}, \bibinfo {author} {\bibfnamefont {P.}~\bibnamefont {Bunyk}}, \bibinfo {author} {\bibfnamefont {E.}~\bibnamefont {Tolkacheva}}, \bibinfo {author} {\bibfnamefont {E.}~\bibnamefont {Ladizinsky}}, \bibinfo {author} {\bibfnamefont {N.}~\bibnamefont {Ladizinsky}}, \bibinfo {author} {\bibfnamefont {T.}~\bibnamefont {Oh}}, \emph {et~al.},\ }\bibfield  {title} {\bibinfo {title} {Experimental investigation of an eight-qubit unit cell in a superconducting optimization processor},\ }\href@noop {} {\bibfield  {journal} {\bibinfo  {journal} {Physical Review B}\ }\textbf {\bibinfo {volume} {82}},\ \bibinfo {pages} {024511} (\bibinfo {year} {2010}{\natexlab{b}})}\BibitemShut {NoStop}%
\bibitem [{\citenamefont {Johnson}\ \emph {et~al.}(2011)\citenamefont {Johnson}, \citenamefont {Amin}, \citenamefont {Gildert}, \citenamefont {Lanting}, \citenamefont {Hamze}, \citenamefont {Dickson}, \citenamefont {Harris}, \citenamefont {Berkley}, \citenamefont {Johansson}, \citenamefont {Bunyk} \emph {et~al.}}]{johnson2011quantum}%
  \BibitemOpen
  \bibfield  {author} {\bibinfo {author} {\bibfnamefont {M.~W.}\ \bibnamefont {Johnson}}, \bibinfo {author} {\bibfnamefont {M.~H.}\ \bibnamefont {Amin}}, \bibinfo {author} {\bibfnamefont {S.}~\bibnamefont {Gildert}}, \bibinfo {author} {\bibfnamefont {T.}~\bibnamefont {Lanting}}, \bibinfo {author} {\bibfnamefont {F.}~\bibnamefont {Hamze}}, \bibinfo {author} {\bibfnamefont {N.}~\bibnamefont {Dickson}}, \bibinfo {author} {\bibfnamefont {R.}~\bibnamefont {Harris}}, \bibinfo {author} {\bibfnamefont {A.~J.}\ \bibnamefont {Berkley}}, \bibinfo {author} {\bibfnamefont {J.}~\bibnamefont {Johansson}}, \bibinfo {author} {\bibfnamefont {P.}~\bibnamefont {Bunyk}}, \emph {et~al.},\ }\bibfield  {title} {\bibinfo {title} {Quantum annealing with manufactured spins},\ }\href@noop {} {\bibfield  {journal} {\bibinfo  {journal} {Nature}\ }\textbf {\bibinfo {volume} {473}},\ \bibinfo {pages} {194} (\bibinfo {year} {2011})}\BibitemShut {NoStop}%
\bibitem [{\citenamefont {King}\ \emph {et~al.}(2023)\citenamefont {King}, \citenamefont {Raymond}, \citenamefont {Lanting}, \citenamefont {Harris}, \citenamefont {Zucca}, \citenamefont {Altomare}, \citenamefont {Berkley}, \citenamefont {Boothby}, \citenamefont {Ejtemaee}, \citenamefont {Enderud} \emph {et~al.}}]{king2023quantum}%
  \BibitemOpen
  \bibfield  {author} {\bibinfo {author} {\bibfnamefont {A.~D.}\ \bibnamefont {King}}, \bibinfo {author} {\bibfnamefont {J.}~\bibnamefont {Raymond}}, \bibinfo {author} {\bibfnamefont {T.}~\bibnamefont {Lanting}}, \bibinfo {author} {\bibfnamefont {R.}~\bibnamefont {Harris}}, \bibinfo {author} {\bibfnamefont {A.}~\bibnamefont {Zucca}}, \bibinfo {author} {\bibfnamefont {F.}~\bibnamefont {Altomare}}, \bibinfo {author} {\bibfnamefont {A.~J.}\ \bibnamefont {Berkley}}, \bibinfo {author} {\bibfnamefont {K.}~\bibnamefont {Boothby}}, \bibinfo {author} {\bibfnamefont {S.}~\bibnamefont {Ejtemaee}}, \bibinfo {author} {\bibfnamefont {C.}~\bibnamefont {Enderud}}, \emph {et~al.},\ }\bibfield  {title} {\bibinfo {title} {Quantum critical dynamics in a 5,000-qubit programmable spin glass},\ }\href@noop {} {\bibfield  {journal} {\bibinfo  {journal} {Nature}\ }\textbf {\bibinfo {volume} {617}},\ \bibinfo {pages} {61} (\bibinfo {year} {2023})}\BibitemShut {NoStop}%
\bibitem [{\citenamefont {Milburn}\ and\ \citenamefont {Holmes}(1991)}]{milburn1991quantum}%
  \BibitemOpen
  \bibfield  {author} {\bibinfo {author} {\bibfnamefont {G.}~\bibnamefont {Milburn}}\ and\ \bibinfo {author} {\bibfnamefont {C.}~\bibnamefont {Holmes}},\ }\bibfield  {title} {\bibinfo {title} {Quantum coherence and classical chaos in a pulsed parametric oscillator with a kerr nonlinearity},\ }\href@noop {} {\bibfield  {journal} {\bibinfo  {journal} {Physical Review A}\ }\textbf {\bibinfo {volume} {44}},\ \bibinfo {pages} {4704} (\bibinfo {year} {1991})}\BibitemShut {NoStop}%
\bibitem [{\citenamefont {Wielinga}\ and\ \citenamefont {Milburn}(1993)}]{wielinga1993quantum}%
  \BibitemOpen
  \bibfield  {author} {\bibinfo {author} {\bibfnamefont {B.}~\bibnamefont {Wielinga}}\ and\ \bibinfo {author} {\bibfnamefont {G.}~\bibnamefont {Milburn}},\ }\bibfield  {title} {\bibinfo {title} {Quantum tunneling in a kerr medium with parametric pumping},\ }\href@noop {} {\bibfield  {journal} {\bibinfo  {journal} {Physical Review A}\ }\textbf {\bibinfo {volume} {48}},\ \bibinfo {pages} {2494} (\bibinfo {year} {1993})}\BibitemShut {NoStop}%
\bibitem [{\citenamefont {Meaney}\ \emph {et~al.}(2014)\citenamefont {Meaney}, \citenamefont {Nha}, \citenamefont {Duty},\ and\ \citenamefont {Milburn}}]{meaney2014quantum}%
  \BibitemOpen
  \bibfield  {author} {\bibinfo {author} {\bibfnamefont {C.~H.}\ \bibnamefont {Meaney}}, \bibinfo {author} {\bibfnamefont {H.}~\bibnamefont {Nha}}, \bibinfo {author} {\bibfnamefont {T.}~\bibnamefont {Duty}},\ and\ \bibinfo {author} {\bibfnamefont {G.~J.}\ \bibnamefont {Milburn}},\ }\bibfield  {title} {\bibinfo {title} {Quantum and classical nonlinear dynamics in a microwave cavity},\ }\href@noop {} {\bibfield  {journal} {\bibinfo  {journal} {EPJ quantum technology}\ }\textbf {\bibinfo {volume} {1}},\ \bibinfo {pages} {1} (\bibinfo {year} {2014})}\BibitemShut {NoStop}%
\bibitem [{\citenamefont {Goto}(2016)}]{goto2016bifurcation}%
  \BibitemOpen
  \bibfield  {author} {\bibinfo {author} {\bibfnamefont {H.}~\bibnamefont {Goto}},\ }\bibfield  {title} {\bibinfo {title} {Bifurcation-based adiabatic quantum computation with a nonlinear oscillator network},\ }\href@noop {} {\bibfield  {journal} {\bibinfo  {journal} {Scientific reports}\ }\textbf {\bibinfo {volume} {6}},\ \bibinfo {pages} {21686} (\bibinfo {year} {2016})}\BibitemShut {NoStop}%
\bibitem [{\citenamefont {Puri}\ \emph {et~al.}(2017{\natexlab{a}})\citenamefont {Puri}, \citenamefont {Boutin},\ and\ \citenamefont {Blais}}]{Puri2017_npjq}%
  \BibitemOpen
  \bibfield  {author} {\bibinfo {author} {\bibfnamefont {S.}~\bibnamefont {Puri}}, \bibinfo {author} {\bibfnamefont {S.}~\bibnamefont {Boutin}},\ and\ \bibinfo {author} {\bibfnamefont {A.}~\bibnamefont {Blais}},\ }\bibfield  {title} {\bibinfo {title} {Engineering the quantum states of light in a kerr-nonlinear resonator by two-photon driving},\ }\href@noop {} {\bibfield  {journal} {\bibinfo  {journal} {npj Quantum Information}\ }\textbf {\bibinfo {volume} {3}},\ \bibinfo {pages} {18} (\bibinfo {year} {2017}{\natexlab{a}})}\BibitemShut {NoStop}%
\bibitem [{\citenamefont {Wang}\ \emph {et~al.}(2019)\citenamefont {Wang}, \citenamefont {Pechal}, \citenamefont {Wollack}, \citenamefont {Arrangoiz-Arriola}, \citenamefont {Gao}, \citenamefont {Lee},\ and\ \citenamefont {Safavi-Naeini}}]{wang2019quantum}%
  \BibitemOpen
  \bibfield  {author} {\bibinfo {author} {\bibfnamefont {Z.}~\bibnamefont {Wang}}, \bibinfo {author} {\bibfnamefont {M.}~\bibnamefont {Pechal}}, \bibinfo {author} {\bibfnamefont {E.~A.}\ \bibnamefont {Wollack}}, \bibinfo {author} {\bibfnamefont {P.}~\bibnamefont {Arrangoiz-Arriola}}, \bibinfo {author} {\bibfnamefont {M.}~\bibnamefont {Gao}}, \bibinfo {author} {\bibfnamefont {N.~R.}\ \bibnamefont {Lee}},\ and\ \bibinfo {author} {\bibfnamefont {A.~H.}\ \bibnamefont {Safavi-Naeini}},\ }\bibfield  {title} {\bibinfo {title} {Quantum dynamics of a few-photon parametric oscillator},\ }\href@noop {} {\bibfield  {journal} {\bibinfo  {journal} {Physical Review X}\ }\textbf {\bibinfo {volume} {9}},\ \bibinfo {pages} {021049} (\bibinfo {year} {2019})}\BibitemShut {NoStop}%
\bibitem [{\citenamefont {Grimm}\ \emph {et~al.}(2020)\citenamefont {Grimm}, \citenamefont {Frattini}, \citenamefont {Puri}, \citenamefont {Mundhada}, \citenamefont {Touzard}, \citenamefont {Mirrahimi}, \citenamefont {Girvin}, \citenamefont {Shankar},\ and\ \citenamefont {Devoret}}]{grimm2020stabilization}%
  \BibitemOpen
  \bibfield  {author} {\bibinfo {author} {\bibfnamefont {A.}~\bibnamefont {Grimm}}, \bibinfo {author} {\bibfnamefont {N.~E.}\ \bibnamefont {Frattini}}, \bibinfo {author} {\bibfnamefont {S.}~\bibnamefont {Puri}}, \bibinfo {author} {\bibfnamefont {S.~O.}\ \bibnamefont {Mundhada}}, \bibinfo {author} {\bibfnamefont {S.}~\bibnamefont {Touzard}}, \bibinfo {author} {\bibfnamefont {M.}~\bibnamefont {Mirrahimi}}, \bibinfo {author} {\bibfnamefont {S.~M.}\ \bibnamefont {Girvin}}, \bibinfo {author} {\bibfnamefont {S.}~\bibnamefont {Shankar}},\ and\ \bibinfo {author} {\bibfnamefont {M.~H.}\ \bibnamefont {Devoret}},\ }\bibfield  {title} {\bibinfo {title} {Stabilization and operation of a kerr-cat qubit},\ }\href@noop {} {\bibfield  {journal} {\bibinfo  {journal} {Nature}\ }\textbf {\bibinfo {volume} {584}},\ \bibinfo {pages} {205} (\bibinfo {year} {2020})}\BibitemShut {NoStop}%
\bibitem [{\citenamefont {Yamaji}\ \emph {et~al.}(2022)\citenamefont {Yamaji}, \citenamefont {Kagami}, \citenamefont {Yamaguchi}, \citenamefont {Satoh}, \citenamefont {Koshino}, \citenamefont {Goto}, \citenamefont {Lin}, \citenamefont {Nakamura},\ and\ \citenamefont {Yamamoto}}]{yamaji2022spectroscopic}%
  \BibitemOpen
  \bibfield  {author} {\bibinfo {author} {\bibfnamefont {T.}~\bibnamefont {Yamaji}}, \bibinfo {author} {\bibfnamefont {S.}~\bibnamefont {Kagami}}, \bibinfo {author} {\bibfnamefont {A.}~\bibnamefont {Yamaguchi}}, \bibinfo {author} {\bibfnamefont {T.}~\bibnamefont {Satoh}}, \bibinfo {author} {\bibfnamefont {K.}~\bibnamefont {Koshino}}, \bibinfo {author} {\bibfnamefont {H.}~\bibnamefont {Goto}}, \bibinfo {author} {\bibfnamefont {Z.}~\bibnamefont {Lin}}, \bibinfo {author} {\bibfnamefont {Y.}~\bibnamefont {Nakamura}},\ and\ \bibinfo {author} {\bibfnamefont {T.}~\bibnamefont {Yamamoto}},\ }\bibfield  {title} {\bibinfo {title} {Spectroscopic observation of the crossover from a classical duffing oscillator to a kerr parametric oscillator},\ }\href@noop {} {\bibfield  {journal} {\bibinfo  {journal} {Physical Review A}\ }\textbf {\bibinfo {volume} {105}},\ \bibinfo {pages} {023519} (\bibinfo {year} {2022})}\BibitemShut {NoStop}%
\bibitem [{\citenamefont {Yamaji}\ \emph {et~al.}(2023)\citenamefont {Yamaji}, \citenamefont {Masuda}, \citenamefont {Yamaguchi}, \citenamefont {Satoh}, \citenamefont {Morioka}, \citenamefont {Igarashi}, \citenamefont {Shirane},\ and\ \citenamefont {Yamamoto}}]{yamaji2023correlated}%
  \BibitemOpen
  \bibfield  {author} {\bibinfo {author} {\bibfnamefont {T.}~\bibnamefont {Yamaji}}, \bibinfo {author} {\bibfnamefont {S.}~\bibnamefont {Masuda}}, \bibinfo {author} {\bibfnamefont {A.}~\bibnamefont {Yamaguchi}}, \bibinfo {author} {\bibfnamefont {T.}~\bibnamefont {Satoh}}, \bibinfo {author} {\bibfnamefont {A.}~\bibnamefont {Morioka}}, \bibinfo {author} {\bibfnamefont {Y.}~\bibnamefont {Igarashi}}, \bibinfo {author} {\bibfnamefont {M.}~\bibnamefont {Shirane}},\ and\ \bibinfo {author} {\bibfnamefont {T.}~\bibnamefont {Yamamoto}},\ }\bibfield  {title} {\bibinfo {title} {Correlated oscillations in kerr parametric oscillators with tunable effective coupling},\ }\href@noop {} {\bibfield  {journal} {\bibinfo  {journal} {Physical Review Applied}\ }\textbf {\bibinfo {volume} {20}},\ \bibinfo {pages} {014057} (\bibinfo {year} {2023})}\BibitemShut {NoStop}%
\bibitem [{\citenamefont {Iyama}\ \emph {et~al.}(2024)\citenamefont {Iyama}, \citenamefont {Kamiya}, \citenamefont {Fujii}, \citenamefont {Mukai}, \citenamefont {Zhou}, \citenamefont {Nagase}, \citenamefont {Tomonaga}, \citenamefont {Wang}, \citenamefont {Xue}, \citenamefont {Watabe} \emph {et~al.}}]{iyama2024observation}%
  \BibitemOpen
  \bibfield  {author} {\bibinfo {author} {\bibfnamefont {D.}~\bibnamefont {Iyama}}, \bibinfo {author} {\bibfnamefont {T.}~\bibnamefont {Kamiya}}, \bibinfo {author} {\bibfnamefont {S.}~\bibnamefont {Fujii}}, \bibinfo {author} {\bibfnamefont {H.}~\bibnamefont {Mukai}}, \bibinfo {author} {\bibfnamefont {Y.}~\bibnamefont {Zhou}}, \bibinfo {author} {\bibfnamefont {T.}~\bibnamefont {Nagase}}, \bibinfo {author} {\bibfnamefont {A.}~\bibnamefont {Tomonaga}}, \bibinfo {author} {\bibfnamefont {R.}~\bibnamefont {Wang}}, \bibinfo {author} {\bibfnamefont {J.-J.}\ \bibnamefont {Xue}}, \bibinfo {author} {\bibfnamefont {S.}~\bibnamefont {Watabe}}, \emph {et~al.},\ }\bibfield  {title} {\bibinfo {title} {Observation and manipulation of quantum interference in a superconducting kerr parametric oscillator},\ }\href@noop {} {\bibfield  {journal} {\bibinfo  {journal} {Nature Communications}\ }\textbf {\bibinfo {volume} {15}},\ \bibinfo {pages} {86} (\bibinfo {year} {2024})}\BibitemShut {NoStop}%
\bibitem [{\citenamefont {Yamaguchi}\ \emph {et~al.}(2023)\citenamefont {Yamaguchi}, \citenamefont {Masuda}, \citenamefont {Matsuzaki}, \citenamefont {Yamaji}, \citenamefont {Satoh}, \citenamefont {Morioka}, \citenamefont {Kawakami}, \citenamefont {Igarashi}, \citenamefont {Shirane},\ and\ \citenamefont {Yamamoto}}]{yamaguchi2023spectroscopy}%
  \BibitemOpen
  \bibfield  {author} {\bibinfo {author} {\bibfnamefont {A.}~\bibnamefont {Yamaguchi}}, \bibinfo {author} {\bibfnamefont {S.}~\bibnamefont {Masuda}}, \bibinfo {author} {\bibfnamefont {Y.}~\bibnamefont {Matsuzaki}}, \bibinfo {author} {\bibfnamefont {T.}~\bibnamefont {Yamaji}}, \bibinfo {author} {\bibfnamefont {T.}~\bibnamefont {Satoh}}, \bibinfo {author} {\bibfnamefont {A.}~\bibnamefont {Morioka}}, \bibinfo {author} {\bibfnamefont {Y.}~\bibnamefont {Kawakami}}, \bibinfo {author} {\bibfnamefont {Y.}~\bibnamefont {Igarashi}}, \bibinfo {author} {\bibfnamefont {M.}~\bibnamefont {Shirane}},\ and\ \bibinfo {author} {\bibfnamefont {T.}~\bibnamefont {Yamamoto}},\ }\bibfield  {title} {\bibinfo {title} {Spectroscopy of flux-driven kerr parametric oscillators by reflection coefficient measurement},\ }\href@noop {} {\bibfield  {journal} {\bibinfo  {journal} {arXiv preprint arXiv:2309.10488}\ } (\bibinfo {year} {2023})}\BibitemShut {NoStop}%
\bibitem [{\citenamefont {Cochrane}\ \emph {et~al.}(2000)\citenamefont {Cochrane}, \citenamefont {Milburn},\ and\ \citenamefont {Munro}}]{cochrane2000teleportation}%
  \BibitemOpen
  \bibfield  {author} {\bibinfo {author} {\bibfnamefont {P.}~\bibnamefont {Cochrane}}, \bibinfo {author} {\bibfnamefont {G.~J.}\ \bibnamefont {Milburn}},\ and\ \bibinfo {author} {\bibfnamefont {W.}~\bibnamefont {Munro}},\ }\bibfield  {title} {\bibinfo {title} {Teleportation using coupled oscillator states},\ }\href@noop {} {\bibfield  {journal} {\bibinfo  {journal} {Physical Review A}\ }\textbf {\bibinfo {volume} {62}},\ \bibinfo {pages} {062307} (\bibinfo {year} {2000})}\BibitemShut {NoStop}%
\bibitem [{\citenamefont {Koch}\ \emph {et~al.}(2007)\citenamefont {Koch}, \citenamefont {Terri}, \citenamefont {Gambetta}, \citenamefont {Houck}, \citenamefont {Schuster}, \citenamefont {Majer}, \citenamefont {Blais}, \citenamefont {Devoret}, \citenamefont {Girvin},\ and\ \citenamefont {Schoelkopf}}]{koch2007charge}%
  \BibitemOpen
  \bibfield  {author} {\bibinfo {author} {\bibfnamefont {J.}~\bibnamefont {Koch}}, \bibinfo {author} {\bibfnamefont {M.~Y.}\ \bibnamefont {Terri}}, \bibinfo {author} {\bibfnamefont {J.}~\bibnamefont {Gambetta}}, \bibinfo {author} {\bibfnamefont {A.~A.}\ \bibnamefont {Houck}}, \bibinfo {author} {\bibfnamefont {D.~I.}\ \bibnamefont {Schuster}}, \bibinfo {author} {\bibfnamefont {J.}~\bibnamefont {Majer}}, \bibinfo {author} {\bibfnamefont {A.}~\bibnamefont {Blais}}, \bibinfo {author} {\bibfnamefont {M.~H.}\ \bibnamefont {Devoret}}, \bibinfo {author} {\bibfnamefont {S.~M.}\ \bibnamefont {Girvin}},\ and\ \bibinfo {author} {\bibfnamefont {R.~J.}\ \bibnamefont {Schoelkopf}},\ }\bibfield  {title} {\bibinfo {title} {Charge-insensitive qubit design derived from the cooper pair box},\ }\href@noop {} {\bibfield  {journal} {\bibinfo  {journal} {Physical Review A}\ }\textbf {\bibinfo {volume} {76}},\ \bibinfo {pages} {042319} (\bibinfo {year} {2007})}\BibitemShut {NoStop}%
\bibitem [{\citenamefont {Schreier}\ \emph {et~al.}(2008)\citenamefont {Schreier}, \citenamefont {Houck}, \citenamefont {Koch}, \citenamefont {Schuster}, \citenamefont {Johnson}, \citenamefont {Chow}, \citenamefont {Gambetta}, \citenamefont {Majer}, \citenamefont {Frunzio}, \citenamefont {Devoret} \emph {et~al.}}]{schreier2008suppressing}%
  \BibitemOpen
  \bibfield  {author} {\bibinfo {author} {\bibfnamefont {J.~A.}\ \bibnamefont {Schreier}}, \bibinfo {author} {\bibfnamefont {A.~A.}\ \bibnamefont {Houck}}, \bibinfo {author} {\bibfnamefont {J.}~\bibnamefont {Koch}}, \bibinfo {author} {\bibfnamefont {D.~I.}\ \bibnamefont {Schuster}}, \bibinfo {author} {\bibfnamefont {B.~R.}\ \bibnamefont {Johnson}}, \bibinfo {author} {\bibfnamefont {J.~M.}\ \bibnamefont {Chow}}, \bibinfo {author} {\bibfnamefont {J.~M.}\ \bibnamefont {Gambetta}}, \bibinfo {author} {\bibfnamefont {J.}~\bibnamefont {Majer}}, \bibinfo {author} {\bibfnamefont {L.}~\bibnamefont {Frunzio}}, \bibinfo {author} {\bibfnamefont {M.~H.}\ \bibnamefont {Devoret}}, \emph {et~al.},\ }\bibfield  {title} {\bibinfo {title} {Suppressing charge noise decoherence in superconducting charge qubits},\ }\href@noop {} {\bibfield  {journal} {\bibinfo  {journal} {Physical Review B}\ }\textbf {\bibinfo {volume} {77}},\ \bibinfo {pages} {180502} (\bibinfo {year} {2008})}\BibitemShut {NoStop}%
\bibitem [{\citenamefont {Santoro}\ \emph {et~al.}(2002)\citenamefont {Santoro}, \citenamefont {Marton{\'a}k}, \citenamefont {Tosatti},\ and\ \citenamefont {Car}}]{santoro2002theory}%
  \BibitemOpen
  \bibfield  {author} {\bibinfo {author} {\bibfnamefont {G.~E.}\ \bibnamefont {Santoro}}, \bibinfo {author} {\bibfnamefont {R.}~\bibnamefont {Marton{\'a}k}}, \bibinfo {author} {\bibfnamefont {E.}~\bibnamefont {Tosatti}},\ and\ \bibinfo {author} {\bibfnamefont {R.}~\bibnamefont {Car}},\ }\bibfield  {title} {\bibinfo {title} {Theory of quantum annealing of an ising spin glass},\ }\href@noop {} {\bibfield  {journal} {\bibinfo  {journal} {Science}\ }\textbf {\bibinfo {volume} {295}},\ \bibinfo {pages} {2427} (\bibinfo {year} {2002})}\BibitemShut {NoStop}%
\bibitem [{\citenamefont {Lechner}\ \emph {et~al.}(2015)\citenamefont {Lechner}, \citenamefont {Hauke},\ and\ \citenamefont {Zoller}}]{lechner2015quantum}%
  \BibitemOpen
  \bibfield  {author} {\bibinfo {author} {\bibfnamefont {W.}~\bibnamefont {Lechner}}, \bibinfo {author} {\bibfnamefont {P.}~\bibnamefont {Hauke}},\ and\ \bibinfo {author} {\bibfnamefont {P.}~\bibnamefont {Zoller}},\ }\bibfield  {title} {\bibinfo {title} {A quantum annealing architecture with all-to-all connectivity from local interactions},\ }\href@noop {} {\bibfield  {journal} {\bibinfo  {journal} {Science advances}\ }\textbf {\bibinfo {volume} {1}},\ \bibinfo {pages} {e1500838} (\bibinfo {year} {2015})}\BibitemShut {NoStop}%
\bibitem [{\citenamefont {Puri}\ \emph {et~al.}(2017{\natexlab{b}})\citenamefont {Puri}, \citenamefont {Andersen}, \citenamefont {Grimsmo},\ and\ \citenamefont {Blais}}]{puri2017quantum}%
  \BibitemOpen
  \bibfield  {author} {\bibinfo {author} {\bibfnamefont {S.}~\bibnamefont {Puri}}, \bibinfo {author} {\bibfnamefont {C.~K.}\ \bibnamefont {Andersen}}, \bibinfo {author} {\bibfnamefont {A.~L.}\ \bibnamefont {Grimsmo}},\ and\ \bibinfo {author} {\bibfnamefont {A.}~\bibnamefont {Blais}},\ }\bibfield  {title} {\bibinfo {title} {Quantum annealing with all-to-all connected nonlinear oscillators},\ }\href@noop {} {\bibfield  {journal} {\bibinfo  {journal} {Nature communications}\ }\textbf {\bibinfo {volume} {8}},\ \bibinfo {pages} {15785} (\bibinfo {year} {2017}{\natexlab{b}})}\BibitemShut {NoStop}%
\bibitem [{\citenamefont {Zhao}\ \emph {et~al.}(2018)\citenamefont {Zhao}, \citenamefont {Jin}, \citenamefont {Xu}, \citenamefont {Tan}, \citenamefont {Yu},\ and\ \citenamefont {Yu}}]{zhao2018two}%
  \BibitemOpen
  \bibfield  {author} {\bibinfo {author} {\bibfnamefont {P.}~\bibnamefont {Zhao}}, \bibinfo {author} {\bibfnamefont {Z.}~\bibnamefont {Jin}}, \bibinfo {author} {\bibfnamefont {P.}~\bibnamefont {Xu}}, \bibinfo {author} {\bibfnamefont {X.}~\bibnamefont {Tan}}, \bibinfo {author} {\bibfnamefont {H.}~\bibnamefont {Yu}},\ and\ \bibinfo {author} {\bibfnamefont {Y.}~\bibnamefont {Yu}},\ }\bibfield  {title} {\bibinfo {title} {Two-photon driven kerr resonator for quantum annealing with three-dimensional circuit qed},\ }\href@noop {} {\bibfield  {journal} {\bibinfo  {journal} {Physical Review Applied}\ }\textbf {\bibinfo {volume} {10}},\ \bibinfo {pages} {024019} (\bibinfo {year} {2018})}\BibitemShut {NoStop}%
\bibitem [{\citenamefont {Razmkhah}\ \emph {et~al.}(2024)\citenamefont {Razmkhah}, \citenamefont {Kamal}, \citenamefont {Yoshikawa},\ and\ \citenamefont {Pedram}}]{razmkhah2024josephson}%
  \BibitemOpen
  \bibfield  {author} {\bibinfo {author} {\bibfnamefont {S.}~\bibnamefont {Razmkhah}}, \bibinfo {author} {\bibfnamefont {M.}~\bibnamefont {Kamal}}, \bibinfo {author} {\bibfnamefont {N.}~\bibnamefont {Yoshikawa}},\ and\ \bibinfo {author} {\bibfnamefont {M.}~\bibnamefont {Pedram}},\ }\bibfield  {title} {\bibinfo {title} {Josephson parametric oscillator based ising machine},\ }\href@noop {} {\bibfield  {journal} {\bibinfo  {journal} {Physical Review B}\ }\textbf {\bibinfo {volume} {109}},\ \bibinfo {pages} {014511} (\bibinfo {year} {2024})}\BibitemShut {NoStop}%
\bibitem [{\citenamefont {Weber}\ \emph {et~al.}(2017)\citenamefont {Weber}, \citenamefont {Samach}, \citenamefont {Hover}, \citenamefont {Gustavsson}, \citenamefont {Kim}, \citenamefont {Melville}, \citenamefont {Rosenberg}, \citenamefont {Sears}, \citenamefont {Yan}, \citenamefont {Yoder} \emph {et~al.}}]{weber2017coherent}%
  \BibitemOpen
  \bibfield  {author} {\bibinfo {author} {\bibfnamefont {S.~J.}\ \bibnamefont {Weber}}, \bibinfo {author} {\bibfnamefont {G.~O.}\ \bibnamefont {Samach}}, \bibinfo {author} {\bibfnamefont {D.}~\bibnamefont {Hover}}, \bibinfo {author} {\bibfnamefont {S.}~\bibnamefont {Gustavsson}}, \bibinfo {author} {\bibfnamefont {D.~K.}\ \bibnamefont {Kim}}, \bibinfo {author} {\bibfnamefont {A.}~\bibnamefont {Melville}}, \bibinfo {author} {\bibfnamefont {D.}~\bibnamefont {Rosenberg}}, \bibinfo {author} {\bibfnamefont {A.~P.}\ \bibnamefont {Sears}}, \bibinfo {author} {\bibfnamefont {F.}~\bibnamefont {Yan}}, \bibinfo {author} {\bibfnamefont {J.~L.}\ \bibnamefont {Yoder}}, \emph {et~al.},\ }\bibfield  {title} {\bibinfo {title} {Coherent coupled qubits for quantum annealing},\ }\href@noop {} {\bibfield  {journal} {\bibinfo  {journal} {Physical Review Applied}\ }\textbf {\bibinfo {volume} {8}},\ \bibinfo {pages} {014004} (\bibinfo {year} {2017})}\BibitemShut {NoStop}%
\bibitem [{\citenamefont {Sakurai}\ and\ \citenamefont {Napolitano}(2020)}]{sakurai2020modern}%
  \BibitemOpen
  \bibfield  {author} {\bibinfo {author} {\bibfnamefont {J.~J.}\ \bibnamefont {Sakurai}}\ and\ \bibinfo {author} {\bibfnamefont {J.}~\bibnamefont {Napolitano}},\ }\href@noop {} {\emph {\bibinfo {title} {Modern quantum mechanics}}}\ (\bibinfo  {publisher} {Cambridge University Press},\ \bibinfo {year} {2020})\BibitemShut {NoStop}%
\end{thebibliography}%

\end{document}